\documentclass[aps,prl,twocolumn,groupedaddress]{revtex4-2}

\usepackage{color}

\usepackage[utf8]{inputenc}
\usepackage[english]{babel}
\usepackage{amsmath,graphicx,enumerate}

\begin{document}


\title{Flocking without alignment interactions in attractive active Brownian particles} 

\author{L. Caprini$^{1}$}
\author{H. L\"owen$^{1}$}
\affiliation{$^1$ Institut f\"{u}r Theoretische Physik II: Weiche Materie, Heinrich-Heine-Universit\"{a}t D\"{u}sseldorf, D-40225 D\"{u}sseldorf, Germany }

\date{\today}


\begin{abstract}
Within a simple model of attractive active Brownian particles, we predict flocking behavior and challenge the widespread idea that alignment interactions are necessary to observe this collective phenomenon.
Here, we show that even non-aligning attractive interactions can lead to a flocking state.
Monitoring the velocity polarization as the order parameter, we reveal the onset of a first-order transition from a disordered phase, characterized by several small clusters, to a flocking phase, where a single flocking cluster is emerging.
The scenario is confirmed by studying the spatial connected correlation function of particle velocities, which reveals scale-free behavior in flocking states and exponential-like decay for non-flocking configurations.
Our predictions can be tested in microscopic and macroscopic experiments showing flocking, such as animals, migrating cells, and active colloids.
\end{abstract}

\maketitle

Several biological and physical systems are, nowadays, classified as ``active'' or ``self-propelled''~\cite{marchetti2013hydrodynamics, elgeti2015physics, bechinger2016active} because of their ability to extract energy from the environment and convert it into directed motion~\cite{gompper20202020}.
They exhibit a plethora of fascinating collective phenomena, starting from the collective motion shown by groups of animals at the macroscopic scale~\cite{cavagna2014bird}:
Fish display schooling in the ocean~\cite{ward2008quorum}, birds flock in the sky~\cite{ballerini2008interaction} while several insects swarm together in large clouds~\cite{cavagna2017dynamic}.
Flocking motion is also typical of inanimate macroscopic systems, such as active granular rods~\cite{kumar2014flocking}.
At the micron scale, similar phenomena have been observed in systems of migrating cells~\cite{alert2020physical}, known as flowing liquids and solids~\cite{malinverno2017endocytic}, as well as in the swarming of highly dense bacteria~\cite{be2020phase} even confined in circular geometry~\cite{liu2021viscoelastic}.
Last but not least, flocking is shown by systems of self-propelled colloids, such as rolling ferromagnetic microparticles~\cite{kaiser2017flocking} and aligning Quincke rollers~\cite{bricard2013emergence, geyer2019freezing}. 

From a theoretical side, the seminal work of Vicsek~\cite{vicsek1995novel} provided a microscopic model suitable to reproduce the flocking phenomena, through a non-equilibrium phase transition characterized by traveling ordered bands~\cite{solon2015phase} and periodic density waves~\cite{caussin2014emergent}.
Alternatively, hydrodynamic theories, originally formulated by Toner and Tu~\cite{toner1995long}, tackle the problem from a hydrodynamic (e.g. macroscopic) perspective~\cite{yang2015hydrodynamics}.
In both cases, the common approach was to explicitly include a mechanism in the microscopic dynamics or in the hydrodynamic equations responsible for the alignment of the particle velocities and the expected collective motion.
Successively, models accounting both for excluded volume effects and effective alignment interactions have been investigated, revealing a rich scenario displaying phase-separation, even characterized by fluid clusters and fast particle turnover~\cite{zhang2021active}, flocking clusters~\cite{chate2008collective, martin2018collective, van2019interrupted, D2SM00385F, pu2017reentrant} and bands~\cite{knezevic2022collective}.
Even if alignment was already codified in the model/theory, Vicsek-like models~\cite{vicsek2012collective} or variants~\cite{chate2008modeling}, such as the inertial spin model~\cite{cavagna2015flocking} or chiral Vicsek models~\cite{liebchen2017collective}, have been successfully employed to show flocking states, and reproduce experiments based on animals~\cite{cavagna2018physics}.

In the absence of alignment interactions, it is known that spherical repulsive active particles are able to show clustering~\cite{palacci2013living} and phase coexistence~\cite{fily2012athermal, buttinoni2013dynamical}, now termed motility induced phase separation (MIPS)~\cite{cates2015motility, gonnella2015motility, marchetti2016minimal} even in the absence of attractive interactions.
This class of collective phenomena differs from that shown by Vicsek models because of the absence of global polar order.
At the first level of comprehension, the interplay between persistent active forces and pure repulsive interactions generates effective attractions~\cite{farage2015effective, wittmann2017effective} between the particles responsible for their aggregation. Further theoretical explanations have been formulated by introducing a modified Maxwell construction~\cite{solon2018generalized, hermann2021phase} for an effective free-energy and supplemented with Cahn-Hilliard equations~\cite{speck2014effective}.
In spite of this equilibrium-like interpretations, MIPS is characterized by a plethora of genuine non-equilibrium properties with no equilibrium counterpart, such as a temperature difference between dense and dilute phases~\cite{mandal2019motility}, negative interfacial tension~\cite{bialke2015negative, hermann2019non, fausti2021capillary}, short-range spatial velocity correlations in the dense phase~\cite{caprini2020spontaneous, caprini2022role}, as well as hexatic phase inside the cluster~\cite{digregorio2018full, digregorio2022unified} and even
micro-phase separation~\cite{caporusso2020motility, shi2020self}.
The dense phase in MIPS does not show global polar order~\cite{marchetti2016minimal, caprini2020hidden}, e.g. does not flock, except if particles have an elongated shape~\cite{grossmann2020particle} or if explicit and implicit alignment interactions are included in the microscopic dynamics~\cite{sese2021phase, lam2015self}.

\begin{figure}[!t]
\centering
\includegraphics[width=0.9\linewidth,keepaspectratio]{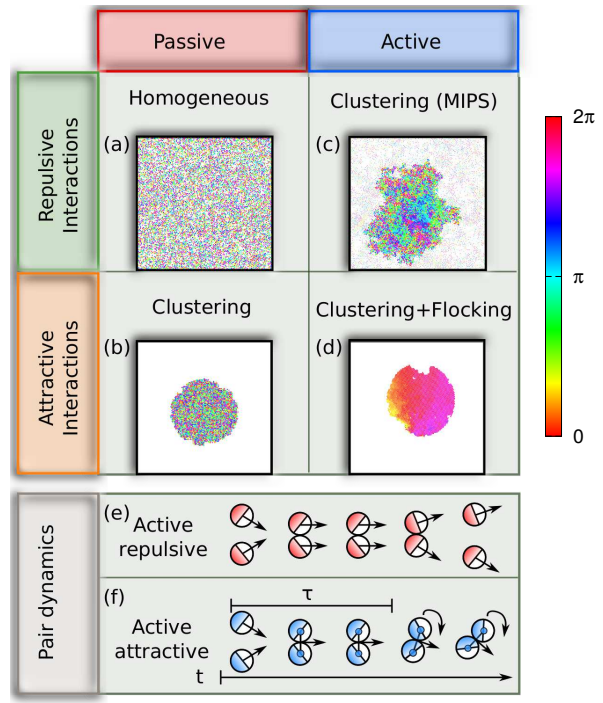}
\caption{Scheme of typical passive and active collective phenomena with repulsive and attractive interactions. In each case, a snapshot configuration of the system is reported, where particles are colored according to their velocity polarization, i.e. the angle formed by their velocity with respect to $x$-axis. 
The first column displays the passive case with repulsive (a) and attractive interactions (b), as a reference.
(a) shows a homogeneous phase while (c) an ordered hexagonal cluster. In both cases, velocities are spatially uncorrelated (random colors). 
The second column shows the active case with repulsive (b) and attractive interactions (d).
(b): The clustering typical of active particles (phase coexistence) is characterized by a vanishing polar order (several colors).
(d): a compact cluster, with a structure similar to that obtained in (c), is reported in this work. The cluster displays a non-vanishing velocity-polar order (same color), i.e. the cluster flocks.
Flocking principle revealed by a collision of two particles: repulsion gives rise to scattering (e), while attraction leads to a stable pair with the same joint velocity (f). 
}\label{fig:Fig1}
\end{figure}

In this Letter, we challenge the widespread idea that flocking behaviors in spherical particles can be observed only in the presence of alignment mechanisms.
As the combination of persistent self-propulsion and pure repulsive interactions generates effective attractions and clustering, we discover that the interplay between persistent active forces and attractive interactions produces strong effective alignment between particles' velocities that can induce a flocking transition. A table with a schematic representation of collective effects in passive and active particles is summarized in Fig.~\ref{fig:Fig1} where the results for both repulsive (Fig.~\ref{fig:Fig1}~(a)-(b)) and attractive (Fig.~\ref{fig:Fig1}~(c)-(d)) interactions are reported.
As a consequence, this Letter shows that the phenomenology of attractive active particles is rather different from that of self-propelled repulsive or passive attractive particles~\cite{binder2021phase}, being characterized by a flocking phenomenon that goes beyond the scenario shown in previous studies~\cite{prymidis2015self, rein2016applicability, paliwal2017non, wachtler2016lane, mallory2017self, mani2015effect} based on coarsening~\cite{alarcon2017morphology, navarro2015clustering} and phase coexistence with reentrant behavior~\cite{redner2013reentrant, prymidis2016vapour, hrishikesh2022collective}.

 
We consider a system of $N$ interacting self-propelled (active) particles in two dimensions, described by underdamped equations of motion for their positions, $\mathbf{x}_i$, and velocities, $\mathbf{v}_i=\dot{\mathbf{x}}_i$, with $i=1, ..., N$.
Each particle is in contact with a thermal bath at temperature $T$ and subject to a friction force $\gamma\mathbf{v}_i$, through the friction coefficient $\gamma$. The active force is included in the dynamics as a stochastic force, $\mathbf{f}^a_i$, which provides to each particle a constant swim velocity, $v_0$, and an orientation vector, $\mathbf{n}_i$, of components $(\cos{\theta_i}, \sin{\theta_i})$.
According to the active Brownian particle (ABP) model~\cite{bechinger2016active, shaebani2020computational}, the orientational angles, $\theta_i$, evolve as independent Brownian processes (no alignment interactions), so that the dynamics reads
\begin{subequations}
\label{eq:wholeABPdynamics}
\begin{align}
\label{eq:v_dynamics}
m\dot{\mathbf{v}}_i &=-\gamma{\mathbf{v}}_i + \mathbf{F}_i + \gamma v_0 \mathbf{n}_i + \sqrt{2T \gamma}\boldsymbol{\eta}_i \\
\label{eq:theta_dynamics}
\dot{\theta}_i&= \sqrt{2D_r} \xi_i \,,
\end{align}
\end{subequations}
where $D_r$ is the rotational diffusion coefficient and $\xi_i$ and $\boldsymbol{\eta}_i$ are white noises with zero average and unit variance.
Particles interact through the force $\mathbf{F}_i=-\nabla_i U_{tot}$,
with $U_{tot} = \sum_{i<j} U(|{\mathbf r}_{ij}|)$ and ${\mathbf r}_{ij}= \mathbf{x}_i -\mathbf{x}_j$.
The shape of the interacting potential $U(r)$ is chosen as an attractive Lennard-Jones potential
$U(r)=4\epsilon\left[\left(\frac{\sigma}{r}\right)^{12}-
\left(\frac{\sigma}{r}\right)^{6}\right]$, for $r\leq
3\sigma$ and zero otherwise.
The constant $\sigma$ represents
the nominal particle diameter while $\epsilon$ is the energy scale of the
interactions.
The system is characterized by two main time scales, the inertial time $\tau_I=m/\gamma$ and the persistence time $\tau=1/D_r$, which determines the time needed by active particles to randomize their orientations.
It is worth mentioning that $v_0 \mathbf{n}_i$ can strongly differ from $\mathbf{v}_i$ in dense configurations.

We consider a box of size $L$ with periodic boundary conditions and integrate the dynamics~\eqref{eq:wholeABPdynamics} with packing fraction $\phi=0.3$ by using an Euler integration scheme. 
Positions and time are rescaled by the nominal particle diameter $\sigma$ and by the persistence time $\tau$, respectively. The resulting dynamics is characterized by several dimensionless parameters: i) the P\'eclet number $\text{\it Pe}=v_0 \tau/\sigma$, quantifying the activity strength that can be also viewed as the ratio between persistence length and particle size, ii) the reduced inertial time, i.e. the ratio between inertial time and persistence time, $\tau_I/\tau$, iii) the translational noise strength $\tau^2T/(m\sigma^2)$, and iv) the reduced potential strength, $\epsilon\tau^2/(m\sigma^2)$.
The latter parameter is set $\gg1$ so that the particles are strongly attractive while the rescaled thermal temperature is chosen $\ll 1$, to neglect the effect of the thermal noise. 
Finally, $\tau_I/\tau \ll 1$ to explore the strongly overdamped case and evaluate large persistence regimes.
In this Letter, we mainly focus on the dependence on $\text{ \it Pe}$ and keep fixed the other dimensionless parameters.
We let the system evolve for a long-time until the coarsening process is achieved and a large cluster containing all the particles is formed as shown in Fig.~\ref{fig:Fig1}~(d).
The bulk of the cluster displays a highly ordered configuration, characterized by an almost perfect hexagonal order similar to that achieved in MIPS with purely repulsive interactions~\cite{redner2013structure}.
At variance with MIPS, where particles leave and join the cluster, attractions make the cluster boundaries more ``stable'' so that particles cannot easily escape and a single dense phase is observed.

In a typical overdamped configuration in the large persistence regime, such that $\tau_I/\tau\ll1$, and moderate $\text{\it Pe}$, the cluster shows flocking behavior despite the absence of any alignment interactions between velocities or self-propulsions.
In other words, even if all the active particles have active forces pointing randomly in space, their velocities are aligned and, as a consequence, the cluster spontaneously displays a net motion. 
This phenomenon is shown in Fig.~\ref{fig:Fig1}~(d), where a snapshot configuration of a flocking cluster is reported. 
Particles are colored according to the direction of their velocity and even a single snapshot shows a directional symmetry breaking (the whole cluster has the same color).
To observe the flocking of clusters, it is crucial that $\tau^2T/(m\sigma^2) \ll 1$ and $\tau_I/\tau \ll 1$, otherwise the phenomenon is suppressed (see SM). 
This flocking cluster is due to the interplay between active forces and attractive interactions,
while it does not occur in repulsive ABP displaying MIPS where the particle velocities in the cluster are only exponentially correlated in space~\cite{caprini2020spontaneous}.
The two different behaviors can be understood by looking at a collision of a particle pair (see also SM):
Pairs of repulsive ABP break when the active force reorients (Fig.~\ref{fig:Fig1}~(e)), while pairs of attractive ABP remain stable thanks to the attractions and show the same joint velocity (Fig.~\ref{fig:Fig1}~(f)).

\begin{figure}[!t]
\centering
\includegraphics[width=1\linewidth,keepaspectratio]{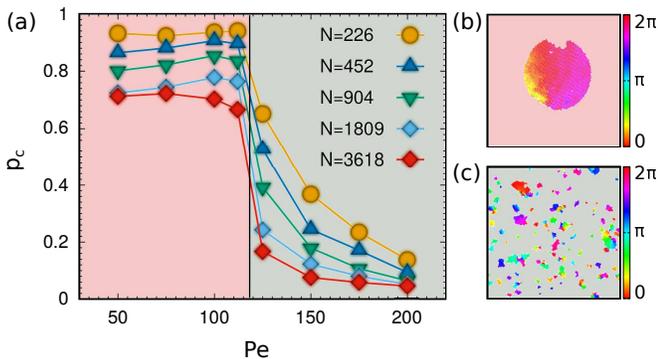}
\caption{Flocking transition. (a): Spatial average polarization, $p_c$, defined in Eq.~\eqref{eq:pc}, as a function of the P\'eclet number, $\text{\it Pe}$, for different system size (number of particles), $N$.
The vertical black line indicates the transition, separating flocking states (pink background) and non-flocking states (grey background) obtained for $\text{\it Pe}_c \approx 2.4\epsilon\tau /(\gamma \sigma^2) \approx 120$.
(b)-(c): Snapshot configurations of the system in the steady state. Particles are colored according to each velocity polarization, i.e. according to the angle formed by each velocity vector with the $x$-direction. (b) and (c) correspond to $\text{\it Pe}=75, 175$, respectively, while $N=1809$ in both panels.
In all cases, errors are smaller than the point size.
The other parameters of the simulations in (a), (b) and (c) are $\tau_I/\tau=10^{-2}$, $\tau^2T/(m\sigma^2)=10^{-3}$, $\epsilon\tau^2/(m\sigma^2)=5 \times 10^3$, and $\phi=0.3$.
}\label{fig:main}
\end{figure}

To quantify the collective motion of the cluster and systematically study the transition towards a flocking state, we consider the velocity polarization as an order parameter, defined as
\begin{equation}
\label{eq:pc}
p_c = \frac{1}{N} \left\langle\left| \sum_{i=0}^{N} \frac{\mathbf{v}_i}{|\mathbf{v}_i|} \right|\right\rangle \,,
\end{equation}
that reads 1 if all the particles move in the same direction and 0 if the directions of the particle velocities are random.
The mean velocity polarization $p_c$ is plotted in Fig.~\ref{fig:main} as a function of $\text{\it Pe}$ and reveals the onset of a first-order phase transition from a flocking state, characterized by $p_c \sim 1$, to a non-flocking state such that $p_c \sim 0$.
This conclusion is supported by our analysis for different system sizes $N$: the larger $N$, the sharper the transition.
For small values of $N$, when $\text{\it Pe}$ is increased, the system is not able to easily reach the non-flocking state and the transition is rather smooth. This occurs because the non-flocking state displays multiple small (and unstable) clusters reminiscent of the ``traveling crystals'' (Fig.~\ref{fig:Fig1}~(c)) observed experimentally~\cite{palacci2013living} and numerically~\cite{mognetti2013living}. Only for large values of $N$, one can observe a sufficiently large number of small clusters (all with different velocity directions) such that $p_c \sim 0$ and the system reaches a vanishing order in the velocity polarization.
The transition line occurs for the value of $\text{\it Pe}$ ($v_0$) needed to overcome the maximal force exerted by neighboring particles, that is $\mathbf{F}_m \approx 2.4 \epsilon$ for our choice of $U(r)$. 
The critical value $\text{\it Pe}_c$ in Fig.~\ref{fig:main}~(a) is calculated by comparing the dimensionless parameters in front of the active force and the maximal force in Eq.~\eqref{eq:v_dynamics}, so that $\text{\it Pe}_c \approx 2.4\epsilon\tau /(\gamma \sigma^2)$ (see also SM).

\begin{figure}[!t]
\centering
\includegraphics[width=0.97\linewidth,keepaspectratio]{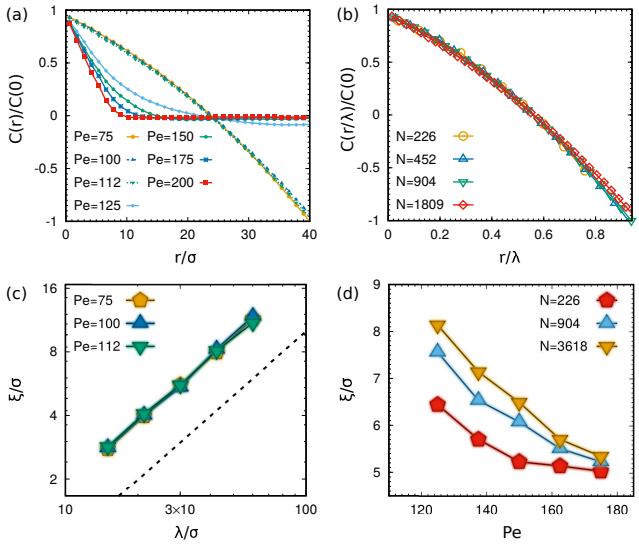}
\caption{Connected correlation functions. (a): connected correlation function, $C(r)/C(0)$, defined in Eq.~\eqref{eq:C(r)}, for different values of the P\'eclet number, $\text{\it Pe}$ (at $N=1809$). (b): $C(r/\lambda)/C(0)$ as a function of the rescaled position $r/\lambda$ where $\lambda=\sigma \sqrt{N}$, for several system size $N$ (at $\text{\it Pe}=100$).
(c) and (d): correlation length $\xi/\sigma$ as a function of $\lambda/\sigma$ (for several values of $\text{\it Pe}$) and $\text{\it Pe}$ (for several values of $N$), respectively. The dotted line in (c) is a guide for the eyes plotting the scaling $\sim \sqrt{N} \sim \lambda$.
In (c), $\xi$ has been calculated as $\xi=r_0/3$, where $r_0$ is the distance such that $C(r=r_0)=0$, while, in (d), as the distance such that $C(r=\xi)=1/e$ being $e$ the Euler constant. Here, errors are smaller than the point size.
The other parameters of the simulations are $\tau_I/\tau=10^{-2}$, $\tau^2T/(m\sigma^2)=10^{-3}$, $\epsilon\tau^2/(m\sigma^2)=5 \times 10^3$, and $\phi=0.3$.
}\label{fig:correlations}
\end{figure}

To confirm the onset of a flocking transition~\cite{cavagna2014bird}, we study the connected correlation function of the velocities, $C(r)$, defined as
\begin{equation}
\label{eq:C(r)}
C(r) = \left\langle\frac{\sum_{i,j}^{N}\delta \mathbf{v}_i \cdot\delta\mathbf{v}_j \delta(r- r_{ij})}{\sum_{i,j}^{N} \delta(r- r_{ij})}\right\rangle \,,
\end{equation}
where $r_{ij}=|\mathbf{x}_i - \mathbf{x}_j|$ represents the distance between particle $i$ and $j$ and $\delta \mathbf{v}_i = \mathbf{v}_i-\sum_{j=0}^N \mathbf{v}_j/N$ measures the fluctuation of $\mathbf{v}_i$ around the spatial average velocity.
The profiles of $C(r)$ are shown in Fig.~\ref{fig:correlations} for several $\text{\it Pe}$ (Fig.~\ref{fig:correlations}~(a)) and system sizes $N$ at fixed $\text{\it Pe}$ (Fig.~\ref{fig:correlations}~(b)).
For $\text{\it Pe}$ corresponding to flocking states, $C(r)$ decays as a power-law, crosses zero, and approaches negative values, as expected in systems showing a flocking transition~\cite{cavagna2018physics}.
In this case, $C(r)$ does not depend on $\text{\it Pe}$ and it is purely determined by the system size, $N$, as reported in Fig.~\ref{fig:correlations}~(b), where the collapse of $C(r)$ for several $\text{\it Pe}$ is shown as the position is rescaled by the cluster size $\lambda \sim\sqrt{N}\sigma$.
In other words, $C(r)$ is scale-free, as expected in flocking configurations and in experiments based on birds~\cite{cavagna2014bird}.
Instead, for values of $\text{\it Pe}$ showing non-flocking states, $C(r)$ has an exponential-like shape and displays a rapid decrease towards zero which becomes faster as $\text{\it Pe}$ is increased (Fig.~\ref{fig:correlations}~(a)).
In this case, the system size plays a marginal role as expected (not shown).

Following Cavagna and Giardina~\cite{cavagna2018physics}, one can define the correlation length, $\xi$, as the distance such that $C(r=\xi)=1/e$ for exponentially decaying $C(r)$ (non-flocking), and as the distance, $\xi=r_0/3$, such that $C(r=r_0)=0$ for algebraically decaying $C(r)$ (flocking states). See also Ref.~\cite{cavagna2022marginal} for further details on this definition.
In Fig.~\ref{fig:correlations}~(c), such a correlation length is shown as a function of the system size $N$ for three values of $\text{\it Pe}$ corresponding to flocking states, confirming that $\xi$ does not depend on $\text{\it Pe}$ and uniquely scales as the cluster size, $\sim\lambda=\sigma\sqrt{N}$.
In Fig.~\ref{fig:correlations}~(d), $\xi$ is plotted as a function of $\text{\it Pe}$ for three different $N$, in the case of non-flocking configurations (large values of $\text{\it Pe}$), where $\xi$ increases as $\text{\it Pe}$ is decreased. 
This occurs because the increase of $\text{\it Pe}$ leads to clusters with smaller sizes: particles belonging to different clusters cannot have correlated velocities.
In this case, the correlation length depends consistently on $N$ only when the system is near the flocking transition: The larger $N$, the larger $\xi$. This behavior for the correlation length is reminiscent of the typical scenario of a first-order phase transition.

To further support the message of the Letter, we derive an analytical prediction that will shed light on the underlying mechanism of the flocking behavior. In the mean-field approximation, we will show that the system is governed by an effective Hamiltonian reminiscent of that introduced for the inertial spin model to describe the behavior of flocks of birds~\cite{cavagna2015flocking}.
At first, we exactly map Eq.~\eqref{eq:wholeABPdynamics} onto a new dynamics by introducing the particle acceleration, $\mathbf{s}_i=\dot{\mathbf{v}}_i$, in the limit of vanishing $T$ - a choice supported by several experiments~\cite{bechinger2016active}. 
We obtain (see Supplemental Material SM)
\begin{subequations}
\label{eq:jerky_dynamics}
\begin{align}
\dot{\mathbf{v}}_i&=\mathbf{s}_i\\
\dot{\mathbf{s}}_i&=\frac{\mathbf{F}_i}{m\tau} -\left(1+\frac{\tau_I}{ \tau}\right)\frac{\mathbf{s}_i}{\tau_I} - \frac{\mathbf{v}_i}{\tau_I \tau} - \frac{\nabla^2_{i j} U}{m}\mathbf{v}_j + \boldsymbol{w}_i 
\end{align}
\end{subequations}
where $\boldsymbol{w}_i=\frac{v_0}{\tau_I}\frac{\sqrt{2}}{\sqrt{\tau}} \boldsymbol{\xi}_i \times \mathbf{n}_i$ and $\boldsymbol{\xi}_i=(0,0,\xi_i)$.
The dynamics~\eqref{eq:jerky_dynamics} could be reminiscent of an inertial spin model~\cite{cavagna2015flocking}. 
Both are characterized by an evolution equation for the second derivative of the velocity and by an effective alignment term $\sim \nabla^2_{ij}U \mathbf{v}_j$ (see SM).
However, the two models do not coincide because the inertial spin model is defined on the lattice and conserves the modulus of the velocity.

To proceed further, we assume that particles are placed in a hexagonal ordered structure (a lattice) because of the strong attracting interactions at play.
This hypothesis is well justified by numerical evidence and by the study of the pair correlation function for instance.
Freezing the particle positions provides a fundamental simplification that allows us to solve the problem under two main additional simplifications.
First, we approximate $\boldsymbol{w}_i$ as a Gaussian white noise, employing the mapping from the ABP dynamics to the active Ornstein-Uhlenbeck model~\cite{maggi2015multidimensional, mandal2017entropy, berthier2019glassy, martin2021statistical, caprini2022parental}, often used to achieve analytical results~\cite{fodor2016far, caprini2018active, das2018confined, dabelow2019irreversibility} in good agreement with ABP simulations~\cite{caprini2020activehigh}.
Secondly, we invoke a mean-field approximation by replacing the coupling between neighboring particles with their average in a two-dimensional hexagonal lattice (see SM).
In this way, we derive analytically the probability distribution of the new dynamics
\begin{equation}
p_m\sim \exp\left(-\mathcal{H}_{f} - \mathcal{H}_{I} \right) \,,
\end{equation}
where $\mathcal{H}_{f}$ and $\mathcal{H}_{I}$ are the effective free (single-particle) and interaction ``Hamiltonians'', respectively, that read
\begin{subequations}
\label{eq:effective_hamiltonians}
\begin{align}
\label{eq:effective_free_H}
\mathcal{H}_{f}&=\sum_i\left[\frac{\tau_I\tau}{v_0^2}\frac{\mathbf{s}_i^2}{2}+\frac{\tau_I\tau}{v_0^2}K \frac{\mathbf{v}_i^2}{2} \right]\,,\\
\label{eq:effective_interacting_H}
\mathcal{H}_{I}&=-\frac{\tau_I\tau}{v_0^2}\frac{K}{6}\sum_{ij}^* \frac{\mathbf{v}_i \cdot \mathbf{v}_j}{2} \,.
\end{align}
\end{subequations}
The sum $\sum^*_{ij}$ is restricted on the first neighbors
and $K=3(U''(\sigma) + U'(\sigma)/\sigma)$ depends on the interacting potential through its derivatives.
$\mathcal{H}_f$ has a Boltzmann shape both for velocities and accelerations variables, which, thus, fluctuate with zero average and an effective (kinetic) temperature given by $T_{eff}=v_0^2/(\tau_I \tau K)$. 
We note that also the inertial spin model allows fluctuation of $\mathbf{s}_i$ (called spin variable in that context) but it does not allow $\mathbf{v}_i$ to fluctuate because of the constraint $|\mathbf{v}_i|=const$.
Finally, the term $\mathcal{H}_I$ provides an alignment effective Hamiltonian reminiscent of that assumed in the theoretical description of the inertial spin model \cite{cavagna2015flocking}.
This term is responsible for the effective alignment interactions observed in the system and is responsible for flocking configurations.

In conclusion, we have shown that alignment interactions are not strictly necessary to achieve flocking in spherical active particles: a minimal and simpler setup to observe flocking clusters is provided by attractive self-propelled particles subject to strongly persistent active forces.
A first-order transition from a non-flocking to a flocking state is achieved through the P\'eclet number, in regimes of large persistence times when thermal fluctuations are small, and is quantitatively supported by the study of the velocity polarization of the system, as an order parameter, and connected correlation functions of the velocity showing scale-free properties.

The contribution of attractions to flocking could be relevant in macroscopic experiments on animals, for instance
midges~\cite{attanasi2014collective} and fly larvae~\cite{dombrovski2017cooperative} that swarm without large aligning interactions,
as well as in microscopic experiments with migrating cell monolayers on a substrate~\cite{malinverno2017endocytic, henkes2020dense} when the polarization in response to forces is small.
In several cases, these systems aggregate forming small-size clusters with effective attractive interactions at play, while the whole cluster could exhibit collective motion similar to the one shown here.
Finally, since strong attractive van-der-Waals forces can occur for colloids that are not index-matched, experiments based on active colloids with persistent self-propulsion~\cite{palacci2013living, buttinoni2013dynamical, van2019interrupted, ginot2018aggregation} represent an ideal platform to verify our predictions.

\begin{acknowledgments}
\textit{Acknowledgments --- } 
We thank Thomas Speck, Benno Liebchen, Antonio Culla, and Andrea Cavagna for helpful discussions.
LC acknowledges support from the Alexander Von Humboldt foundation.
HL acknowledge support by the Deutsche Forschungsgemeinschaft (DFG) through the SPP 2265 under the grant number LO 418/25-1.
\end{acknowledgments}



\bibliographystyle{apsrev4-1}

\bibliography{bib.bib}

\begin{thebibliography}{88}%
\makeatletter
\providecommand \@ifxundefined [1]{%
 \@ifx{#1\undefined}
}%
\providecommand \@ifnum [1]{%
 \ifnum #1\expandafter \@firstoftwo
 \else \expandafter \@secondoftwo
 \fi
}%
\providecommand \@ifx [1]{%
 \ifx #1\expandafter \@firstoftwo
 \else \expandafter \@secondoftwo
 \fi
}%
\providecommand \natexlab [1]{#1}%
\providecommand \enquote  [1]{``#1''}%
\providecommand \bibnamefont  [1]{#1}%
\providecommand \bibfnamefont [1]{#1}%
\providecommand \citenamefont [1]{#1}%
\providecommand \href@noop [0]{\@secondoftwo}%
\providecommand \href [0]{\begingroup \@sanitize@url \@href}%
\providecommand \@href[1]{\@@startlink{#1}\@@href}%
\providecommand \@@href[1]{\endgroup#1\@@endlink}%
\providecommand \@sanitize@url [0]{\catcode `\\12\catcode `\$12\catcode
  `\&12\catcode `\#12\catcode `\^12\catcode `\_12\catcode `\%12\relax}%
\providecommand \@@startlink[1]{}%
\providecommand \@@endlink[0]{}%
\providecommand \url  [0]{\begingroup\@sanitize@url \@url }%
\providecommand \@url [1]{\endgroup\@href {#1}{\urlprefix }}%
\providecommand \urlprefix  [0]{URL }%
\providecommand \Eprint [0]{\href }%
\providecommand \doibase [0]{http://dx.doi.org/}%
\providecommand \selectlanguage [0]{\@gobble}%
\providecommand \bibinfo  [0]{\@secondoftwo}%
\providecommand \bibfield  [0]{\@secondoftwo}%
\providecommand \translation [1]{[#1]}%
\providecommand \BibitemOpen [0]{}%
\providecommand \bibitemStop [0]{}%
\providecommand \bibitemNoStop [0]{.\EOS\space}%
\providecommand \EOS [0]{\spacefactor3000\relax}%
\providecommand \BibitemShut  [1]{\csname bibitem#1\endcsname}%
\let\auto@bib@innerbib\@empty
\bibitem [{\citenamefont {Marchetti}\ \emph {et~al.}(2013)\citenamefont
  {Marchetti}, \citenamefont {Joanny}, \citenamefont {Ramaswamy}, \citenamefont
  {Liverpool}, \citenamefont {Prost}, \citenamefont {Rao},\ and\ \citenamefont
  {Simha}}]{marchetti2013hydrodynamics}%
  \BibitemOpen
  \bibfield  {author} {\bibinfo {author} {\bibfnamefont {M.}~\bibnamefont
  {Marchetti}}, \bibinfo {author} {\bibfnamefont {J.}~\bibnamefont {Joanny}},
  \bibinfo {author} {\bibfnamefont {S.}~\bibnamefont {Ramaswamy}}, \bibinfo
  {author} {\bibfnamefont {T.}~\bibnamefont {Liverpool}}, \bibinfo {author}
  {\bibfnamefont {J.}~\bibnamefont {Prost}}, \bibinfo {author} {\bibfnamefont
  {M.}~\bibnamefont {Rao}}, \ and\ \bibinfo {author} {\bibfnamefont {R.~A.}\
  \bibnamefont {Simha}},\ }\href@noop {} {\bibfield  {journal} {\bibinfo
  {journal} {Rev. Mod. Phys.}\ }\textbf {\bibinfo {volume} {85}},\ \bibinfo
  {pages} {1143} (\bibinfo {year} {2013})}\BibitemShut {NoStop}%
\bibitem [{\citenamefont {Elgeti}\ \emph {et~al.}(2015)\citenamefont {Elgeti},
  \citenamefont {Winkler},\ and\ \citenamefont {Gompper}}]{elgeti2015physics}%
  \BibitemOpen
  \bibfield  {author} {\bibinfo {author} {\bibfnamefont {J.}~\bibnamefont
  {Elgeti}}, \bibinfo {author} {\bibfnamefont {R.~G.}\ \bibnamefont {Winkler}},
  \ and\ \bibinfo {author} {\bibfnamefont {G.}~\bibnamefont {Gompper}},\
  }\href@noop {} {\bibfield  {journal} {\bibinfo  {journal} {Rep. Prog. Phys.}\
  }\textbf {\bibinfo {volume} {78}},\ \bibinfo {pages} {056601} (\bibinfo
  {year} {2015})}\BibitemShut {NoStop}%
\bibitem [{\citenamefont {Bechinger}\ \emph {et~al.}(2016)\citenamefont
  {Bechinger}, \citenamefont {Di~Leonardo}, \citenamefont {L{\"o}wen},
  \citenamefont {Reichhardt}, \citenamefont {Volpe},\ and\ \citenamefont
  {Volpe}}]{bechinger2016active}%
  \BibitemOpen
  \bibfield  {author} {\bibinfo {author} {\bibfnamefont {C.}~\bibnamefont
  {Bechinger}}, \bibinfo {author} {\bibfnamefont {R.}~\bibnamefont
  {Di~Leonardo}}, \bibinfo {author} {\bibfnamefont {H.}~\bibnamefont
  {L{\"o}wen}}, \bibinfo {author} {\bibfnamefont {C.}~\bibnamefont
  {Reichhardt}}, \bibinfo {author} {\bibfnamefont {G.}~\bibnamefont {Volpe}}, \
  and\ \bibinfo {author} {\bibfnamefont {G.}~\bibnamefont {Volpe}},\
  }\href@noop {} {\bibfield  {journal} {\bibinfo  {journal} {Rev. Mod. Phys.}\
  }\textbf {\bibinfo {volume} {88}},\ \bibinfo {pages} {045006} (\bibinfo
  {year} {2016})}\BibitemShut {NoStop}%
\bibitem [{\citenamefont {Gompper}\ \emph {et~al.}(2020)\citenamefont
  {Gompper}, \citenamefont {Winkler}, \citenamefont {Speck}, \citenamefont
  {Solon}, \citenamefont {Nardini}, \citenamefont {Peruani}, \citenamefont
  {L{\"o}wen}, \citenamefont {Golestanian}, \citenamefont {Kaupp},
  \citenamefont {Alvarez} \emph {et~al.}}]{gompper20202020}%
  \BibitemOpen
  \bibfield  {author} {\bibinfo {author} {\bibfnamefont {G.}~\bibnamefont
  {Gompper}}, \bibinfo {author} {\bibfnamefont {R.~G.}\ \bibnamefont
  {Winkler}}, \bibinfo {author} {\bibfnamefont {T.}~\bibnamefont {Speck}},
  \bibinfo {author} {\bibfnamefont {A.}~\bibnamefont {Solon}}, \bibinfo
  {author} {\bibfnamefont {C.}~\bibnamefont {Nardini}}, \bibinfo {author}
  {\bibfnamefont {F.}~\bibnamefont {Peruani}}, \bibinfo {author} {\bibfnamefont
  {H.}~\bibnamefont {L{\"o}wen}}, \bibinfo {author} {\bibfnamefont
  {R.}~\bibnamefont {Golestanian}}, \bibinfo {author} {\bibfnamefont {U.~B.}\
  \bibnamefont {Kaupp}}, \bibinfo {author} {\bibfnamefont {L.}~\bibnamefont
  {Alvarez}},  \emph {et~al.},\ }\href@noop {} {\bibfield  {journal} {\bibinfo
  {journal} {J. Phys. Condens.}\ }\textbf {\bibinfo {volume} {32}},\ \bibinfo
  {pages} {193001} (\bibinfo {year} {2020})}\BibitemShut {NoStop}%
\bibitem [{\citenamefont {Cavagna}\ and\ \citenamefont
  {Giardina}(2014)}]{cavagna2014bird}%
  \BibitemOpen
  \bibfield  {author} {\bibinfo {author} {\bibfnamefont {A.}~\bibnamefont
  {Cavagna}}\ and\ \bibinfo {author} {\bibfnamefont {I.}~\bibnamefont
  {Giardina}},\ }\href@noop {} {\bibfield  {journal} {\bibinfo  {journal}
  {Annu. Rev. Condens. Matter Phys.}\ }\textbf {\bibinfo {volume} {5}},\
  \bibinfo {pages} {183} (\bibinfo {year} {2014})}\BibitemShut {NoStop}%
\bibitem [{\citenamefont {Ward}\ \emph {et~al.}(2008)\citenamefont {Ward},
  \citenamefont {Sumpter}, \citenamefont {Couzin}, \citenamefont {Hart},\ and\
  \citenamefont {Krause}}]{ward2008quorum}%
  \BibitemOpen
  \bibfield  {author} {\bibinfo {author} {\bibfnamefont {A.~J.}\ \bibnamefont
  {Ward}}, \bibinfo {author} {\bibfnamefont {D.~J.}\ \bibnamefont {Sumpter}},
  \bibinfo {author} {\bibfnamefont {I.~D.}\ \bibnamefont {Couzin}}, \bibinfo
  {author} {\bibfnamefont {P.~J.}\ \bibnamefont {Hart}}, \ and\ \bibinfo
  {author} {\bibfnamefont {J.}~\bibnamefont {Krause}},\ }\href@noop {}
  {\bibfield  {journal} {\bibinfo  {journal} {Proc. Natl. Acad. Sci.}\ }\textbf
  {\bibinfo {volume} {105}},\ \bibinfo {pages} {6948} (\bibinfo {year}
  {2008})}\BibitemShut {NoStop}%
\bibitem [{\citenamefont {Ballerini}\ \emph {et~al.}(2008)\citenamefont
  {Ballerini}, \citenamefont {Cabibbo}, \citenamefont {Candelier},
  \citenamefont {Cavagna}, \citenamefont {Cisbani}, \citenamefont {Giardina},
  \citenamefont {Lecomte}, \citenamefont {Orlandi}, \citenamefont {Parisi},
  \citenamefont {Procaccini} \emph {et~al.}}]{ballerini2008interaction}%
  \BibitemOpen
  \bibfield  {author} {\bibinfo {author} {\bibfnamefont {M.}~\bibnamefont
  {Ballerini}}, \bibinfo {author} {\bibfnamefont {N.}~\bibnamefont {Cabibbo}},
  \bibinfo {author} {\bibfnamefont {R.}~\bibnamefont {Candelier}}, \bibinfo
  {author} {\bibfnamefont {A.}~\bibnamefont {Cavagna}}, \bibinfo {author}
  {\bibfnamefont {E.}~\bibnamefont {Cisbani}}, \bibinfo {author} {\bibfnamefont
  {I.}~\bibnamefont {Giardina}}, \bibinfo {author} {\bibfnamefont
  {V.}~\bibnamefont {Lecomte}}, \bibinfo {author} {\bibfnamefont
  {A.}~\bibnamefont {Orlandi}}, \bibinfo {author} {\bibfnamefont
  {G.}~\bibnamefont {Parisi}}, \bibinfo {author} {\bibfnamefont
  {A.}~\bibnamefont {Procaccini}},  \emph {et~al.},\ }\href@noop {} {\bibfield
  {journal} {\bibinfo  {journal} {Proc. Natl. Acad. Sci.}\ }\textbf {\bibinfo
  {volume} {105}},\ \bibinfo {pages} {1232} (\bibinfo {year}
  {2008})}\BibitemShut {NoStop}%
\bibitem [{\citenamefont {Cavagna}\ \emph {et~al.}(2017)\citenamefont
  {Cavagna}, \citenamefont {Conti}, \citenamefont {Creato}, \citenamefont
  {Del~Castello}, \citenamefont {Giardina}, \citenamefont {Grigera},
  \citenamefont {Melillo}, \citenamefont {Parisi},\ and\ \citenamefont
  {Viale}}]{cavagna2017dynamic}%
  \BibitemOpen
  \bibfield  {author} {\bibinfo {author} {\bibfnamefont {A.}~\bibnamefont
  {Cavagna}}, \bibinfo {author} {\bibfnamefont {D.}~\bibnamefont {Conti}},
  \bibinfo {author} {\bibfnamefont {C.}~\bibnamefont {Creato}}, \bibinfo
  {author} {\bibfnamefont {L.}~\bibnamefont {Del~Castello}}, \bibinfo {author}
  {\bibfnamefont {I.}~\bibnamefont {Giardina}}, \bibinfo {author}
  {\bibfnamefont {T.~S.}\ \bibnamefont {Grigera}}, \bibinfo {author}
  {\bibfnamefont {S.}~\bibnamefont {Melillo}}, \bibinfo {author} {\bibfnamefont
  {L.}~\bibnamefont {Parisi}}, \ and\ \bibinfo {author} {\bibfnamefont
  {M.}~\bibnamefont {Viale}},\ }\href@noop {} {\bibfield  {journal} {\bibinfo
  {journal} {Nat. Phys.}\ }\textbf {\bibinfo {volume} {13}},\ \bibinfo {pages}
  {914} (\bibinfo {year} {2017})}\BibitemShut {NoStop}%
\bibitem [{\citenamefont {Kumar}\ \emph {et~al.}(2014)\citenamefont {Kumar},
  \citenamefont {Soni}, \citenamefont {Ramaswamy},\ and\ \citenamefont
  {Sood}}]{kumar2014flocking}%
  \BibitemOpen
  \bibfield  {author} {\bibinfo {author} {\bibfnamefont {N.}~\bibnamefont
  {Kumar}}, \bibinfo {author} {\bibfnamefont {H.}~\bibnamefont {Soni}},
  \bibinfo {author} {\bibfnamefont {S.}~\bibnamefont {Ramaswamy}}, \ and\
  \bibinfo {author} {\bibfnamefont {A.}~\bibnamefont {Sood}},\ }\href@noop {}
  {\bibfield  {journal} {\bibinfo  {journal} {Nat. Commun.}\ }\textbf {\bibinfo
  {volume} {5}},\ \bibinfo {pages} {4688} (\bibinfo {year} {2014})}\BibitemShut
  {NoStop}%
\bibitem [{\citenamefont {Alert}\ and\ \citenamefont
  {Trepat}(2020)}]{alert2020physical}%
  \BibitemOpen
  \bibfield  {author} {\bibinfo {author} {\bibfnamefont {R.}~\bibnamefont
  {Alert}}\ and\ \bibinfo {author} {\bibfnamefont {X.}~\bibnamefont {Trepat}},\
  }\href@noop {} {\bibfield  {journal} {\bibinfo  {journal} {Annu. Rev.
  Condens. Matter Phys.}\ }\textbf {\bibinfo {volume} {11}},\ \bibinfo {pages}
  {77} (\bibinfo {year} {2020})}\BibitemShut {NoStop}%
\bibitem [{\citenamefont {Malinverno}\ \emph {et~al.}(2017)\citenamefont
  {Malinverno}, \citenamefont {Corallino}, \citenamefont {Giavazzi},
  \citenamefont {Bergert}, \citenamefont {Li}, \citenamefont {Leoni},
  \citenamefont {Disanza}, \citenamefont {Frittoli}, \citenamefont {Oldani},
  \citenamefont {Martini} \emph {et~al.}}]{malinverno2017endocytic}%
  \BibitemOpen
  \bibfield  {author} {\bibinfo {author} {\bibfnamefont {C.}~\bibnamefont
  {Malinverno}}, \bibinfo {author} {\bibfnamefont {S.}~\bibnamefont
  {Corallino}}, \bibinfo {author} {\bibfnamefont {F.}~\bibnamefont {Giavazzi}},
  \bibinfo {author} {\bibfnamefont {M.}~\bibnamefont {Bergert}}, \bibinfo
  {author} {\bibfnamefont {Q.}~\bibnamefont {Li}}, \bibinfo {author}
  {\bibfnamefont {M.}~\bibnamefont {Leoni}}, \bibinfo {author} {\bibfnamefont
  {A.}~\bibnamefont {Disanza}}, \bibinfo {author} {\bibfnamefont
  {E.}~\bibnamefont {Frittoli}}, \bibinfo {author} {\bibfnamefont
  {A.}~\bibnamefont {Oldani}}, \bibinfo {author} {\bibfnamefont
  {E.}~\bibnamefont {Martini}},  \emph {et~al.},\ }\href@noop {} {\bibfield
  {journal} {\bibinfo  {journal} {Nat. Mater.}\ }\textbf {\bibinfo {volume}
  {16}},\ \bibinfo {pages} {587} (\bibinfo {year} {2017})}\BibitemShut
  {NoStop}%
\bibitem [{\citenamefont {Be’er}\ \emph {et~al.}(2020)\citenamefont
  {Be’er}, \citenamefont {Ilkanaiv}, \citenamefont {Gross}, \citenamefont
  {Kearns}, \citenamefont {Heidenreich}, \citenamefont {B{\"a}r},\ and\
  \citenamefont {Ariel}}]{be2020phase}%
  \BibitemOpen
  \bibfield  {author} {\bibinfo {author} {\bibfnamefont {A.}~\bibnamefont
  {Be’er}}, \bibinfo {author} {\bibfnamefont {B.}~\bibnamefont {Ilkanaiv}},
  \bibinfo {author} {\bibfnamefont {R.}~\bibnamefont {Gross}}, \bibinfo
  {author} {\bibfnamefont {D.~B.}\ \bibnamefont {Kearns}}, \bibinfo {author}
  {\bibfnamefont {S.}~\bibnamefont {Heidenreich}}, \bibinfo {author}
  {\bibfnamefont {M.}~\bibnamefont {B{\"a}r}}, \ and\ \bibinfo {author}
  {\bibfnamefont {G.}~\bibnamefont {Ariel}},\ }\href@noop {} {\bibfield
  {journal} {\bibinfo  {journal} {Commun. Phys.}\ }\textbf {\bibinfo {volume}
  {3}},\ \bibinfo {pages} {66} (\bibinfo {year} {2020})}\BibitemShut {NoStop}%
\bibitem [{\citenamefont {Liu}\ \emph {et~al.}(2021)\citenamefont {Liu},
  \citenamefont {Shankar}, \citenamefont {Marchetti},\ and\ \citenamefont
  {Wu}}]{liu2021viscoelastic}%
  \BibitemOpen
  \bibfield  {author} {\bibinfo {author} {\bibfnamefont {S.}~\bibnamefont
  {Liu}}, \bibinfo {author} {\bibfnamefont {S.}~\bibnamefont {Shankar}},
  \bibinfo {author} {\bibfnamefont {M.~C.}\ \bibnamefont {Marchetti}}, \ and\
  \bibinfo {author} {\bibfnamefont {Y.}~\bibnamefont {Wu}},\ }\href@noop {}
  {\bibfield  {journal} {\bibinfo  {journal} {Nat.}\ }\textbf {\bibinfo
  {volume} {590}},\ \bibinfo {pages} {80} (\bibinfo {year} {2021})}\BibitemShut
  {NoStop}%
\bibitem [{\citenamefont {Kaiser}\ \emph {et~al.}(2017)\citenamefont {Kaiser},
  \citenamefont {Snezhko},\ and\ \citenamefont {Aranson}}]{kaiser2017flocking}%
  \BibitemOpen
  \bibfield  {author} {\bibinfo {author} {\bibfnamefont {A.}~\bibnamefont
  {Kaiser}}, \bibinfo {author} {\bibfnamefont {A.}~\bibnamefont {Snezhko}}, \
  and\ \bibinfo {author} {\bibfnamefont {I.~S.}\ \bibnamefont {Aranson}},\
  }\href@noop {} {\bibfield  {journal} {\bibinfo  {journal} {Sci. Adv.}\
  }\textbf {\bibinfo {volume} {3}},\ \bibinfo {pages} {e1601469} (\bibinfo
  {year} {2017})}\BibitemShut {NoStop}%
\bibitem [{\citenamefont {Bricard}\ \emph {et~al.}(2013)\citenamefont
  {Bricard}, \citenamefont {Caussin}, \citenamefont {Desreumaux}, \citenamefont
  {Dauchot},\ and\ \citenamefont {Bartolo}}]{bricard2013emergence}%
  \BibitemOpen
  \bibfield  {author} {\bibinfo {author} {\bibfnamefont {A.}~\bibnamefont
  {Bricard}}, \bibinfo {author} {\bibfnamefont {J.-B.}\ \bibnamefont
  {Caussin}}, \bibinfo {author} {\bibfnamefont {N.}~\bibnamefont {Desreumaux}},
  \bibinfo {author} {\bibfnamefont {O.}~\bibnamefont {Dauchot}}, \ and\
  \bibinfo {author} {\bibfnamefont {D.}~\bibnamefont {Bartolo}},\ }\href@noop
  {} {\bibfield  {journal} {\bibinfo  {journal} {Nat.}\ }\textbf {\bibinfo
  {volume} {503}},\ \bibinfo {pages} {95} (\bibinfo {year} {2013})}\BibitemShut
  {NoStop}%
\bibitem [{\citenamefont {Geyer}\ \emph {et~al.}(2019)\citenamefont {Geyer},
  \citenamefont {Martin}, \citenamefont {Tailleur},\ and\ \citenamefont
  {Bartolo}}]{geyer2019freezing}%
  \BibitemOpen
  \bibfield  {author} {\bibinfo {author} {\bibfnamefont {D.}~\bibnamefont
  {Geyer}}, \bibinfo {author} {\bibfnamefont {D.}~\bibnamefont {Martin}},
  \bibinfo {author} {\bibfnamefont {J.}~\bibnamefont {Tailleur}}, \ and\
  \bibinfo {author} {\bibfnamefont {D.}~\bibnamefont {Bartolo}},\ }\href@noop
  {} {\bibfield  {journal} {\bibinfo  {journal} {Phys. Rev. X}\ }\textbf
  {\bibinfo {volume} {9}},\ \bibinfo {pages} {031043} (\bibinfo {year}
  {2019})}\BibitemShut {NoStop}%
\bibitem [{\citenamefont {Vicsek}\ \emph {et~al.}(1995)\citenamefont {Vicsek},
  \citenamefont {Czir{\'o}k}, \citenamefont {Ben-Jacob}, \citenamefont
  {Cohen},\ and\ \citenamefont {Shochet}}]{vicsek1995novel}%
  \BibitemOpen
  \bibfield  {author} {\bibinfo {author} {\bibfnamefont {T.}~\bibnamefont
  {Vicsek}}, \bibinfo {author} {\bibfnamefont {A.}~\bibnamefont {Czir{\'o}k}},
  \bibinfo {author} {\bibfnamefont {E.}~\bibnamefont {Ben-Jacob}}, \bibinfo
  {author} {\bibfnamefont {I.}~\bibnamefont {Cohen}}, \ and\ \bibinfo {author}
  {\bibfnamefont {O.}~\bibnamefont {Shochet}},\ }\href@noop {} {\bibfield
  {journal} {\bibinfo  {journal} {Phys. Rev. Lett.}\ }\textbf {\bibinfo
  {volume} {75}},\ \bibinfo {pages} {1226} (\bibinfo {year}
  {1995})}\BibitemShut {NoStop}%
\bibitem [{\citenamefont {Solon}\ \emph {et~al.}(2015)\citenamefont {Solon},
  \citenamefont {Chat{\'e}},\ and\ \citenamefont {Tailleur}}]{solon2015phase}%
  \BibitemOpen
  \bibfield  {author} {\bibinfo {author} {\bibfnamefont {A.~P.}\ \bibnamefont
  {Solon}}, \bibinfo {author} {\bibfnamefont {H.}~\bibnamefont {Chat{\'e}}}, \
  and\ \bibinfo {author} {\bibfnamefont {J.}~\bibnamefont {Tailleur}},\
  }\href@noop {} {\bibfield  {journal} {\bibinfo  {journal} {Phys. Rev. Lett.}\
  }\textbf {\bibinfo {volume} {114}},\ \bibinfo {pages} {068101} (\bibinfo
  {year} {2015})}\BibitemShut {NoStop}%
\bibitem [{\citenamefont {Caussin}\ \emph {et~al.}(2014)\citenamefont
  {Caussin}, \citenamefont {Solon}, \citenamefont {Peshkov}, \citenamefont
  {Chat{\'e}}, \citenamefont {Dauxois}, \citenamefont {Tailleur}, \citenamefont
  {Vitelli},\ and\ \citenamefont {Bartolo}}]{caussin2014emergent}%
  \BibitemOpen
  \bibfield  {author} {\bibinfo {author} {\bibfnamefont {J.-B.}\ \bibnamefont
  {Caussin}}, \bibinfo {author} {\bibfnamefont {A.}~\bibnamefont {Solon}},
  \bibinfo {author} {\bibfnamefont {A.}~\bibnamefont {Peshkov}}, \bibinfo
  {author} {\bibfnamefont {H.}~\bibnamefont {Chat{\'e}}}, \bibinfo {author}
  {\bibfnamefont {T.}~\bibnamefont {Dauxois}}, \bibinfo {author} {\bibfnamefont
  {J.}~\bibnamefont {Tailleur}}, \bibinfo {author} {\bibfnamefont
  {V.}~\bibnamefont {Vitelli}}, \ and\ \bibinfo {author} {\bibfnamefont
  {D.}~\bibnamefont {Bartolo}},\ }\href@noop {} {\bibfield  {journal} {\bibinfo
   {journal} {Phys. Rev. Lett.}\ }\textbf {\bibinfo {volume} {112}},\ \bibinfo
  {pages} {148102} (\bibinfo {year} {2014})}\BibitemShut {NoStop}%
\bibitem [{\citenamefont {Toner}\ and\ \citenamefont
  {Tu}(1995)}]{toner1995long}%
  \BibitemOpen
  \bibfield  {author} {\bibinfo {author} {\bibfnamefont {J.}~\bibnamefont
  {Toner}}\ and\ \bibinfo {author} {\bibfnamefont {Y.}~\bibnamefont {Tu}},\
  }\href@noop {} {\bibfield  {journal} {\bibinfo  {journal} {Phys. Rev. Lett.}\
  }\textbf {\bibinfo {volume} {75}},\ \bibinfo {pages} {4326} (\bibinfo {year}
  {1995})}\BibitemShut {NoStop}%
\bibitem [{\citenamefont {Yang}\ and\ \citenamefont
  {Marchetti}(2015)}]{yang2015hydrodynamics}%
  \BibitemOpen
  \bibfield  {author} {\bibinfo {author} {\bibfnamefont {X.}~\bibnamefont
  {Yang}}\ and\ \bibinfo {author} {\bibfnamefont {M.~C.}\ \bibnamefont
  {Marchetti}},\ }\href@noop {} {\bibfield  {journal} {\bibinfo  {journal}
  {Phys. Rev. Lett.}\ }\textbf {\bibinfo {volume} {115}},\ \bibinfo {pages}
  {258101} (\bibinfo {year} {2015})}\BibitemShut {NoStop}%
\bibitem [{\citenamefont {Zhang}\ \emph {et~al.}(2021)\citenamefont {Zhang},
  \citenamefont {Alert}, \citenamefont {Yan}, \citenamefont {Wingreen},\ and\
  \citenamefont {Granick}}]{zhang2021active}%
  \BibitemOpen
  \bibfield  {author} {\bibinfo {author} {\bibfnamefont {J.}~\bibnamefont
  {Zhang}}, \bibinfo {author} {\bibfnamefont {R.}~\bibnamefont {Alert}},
  \bibinfo {author} {\bibfnamefont {J.}~\bibnamefont {Yan}}, \bibinfo {author}
  {\bibfnamefont {N.~S.}\ \bibnamefont {Wingreen}}, \ and\ \bibinfo {author}
  {\bibfnamefont {S.}~\bibnamefont {Granick}},\ }\href@noop {} {\bibfield
  {journal} {\bibinfo  {journal} {Nat. Phys.}\ }\textbf {\bibinfo {volume}
  {17}},\ \bibinfo {pages} {961} (\bibinfo {year} {2021})}\BibitemShut
  {NoStop}%
\bibitem [{\citenamefont {Chat{\'e}}\ \emph
  {et~al.}(2008{\natexlab{a}})\citenamefont {Chat{\'e}}, \citenamefont
  {Ginelli}, \citenamefont {Gr{\'e}goire},\ and\ \citenamefont
  {Raynaud}}]{chate2008collective}%
  \BibitemOpen
  \bibfield  {author} {\bibinfo {author} {\bibfnamefont {H.}~\bibnamefont
  {Chat{\'e}}}, \bibinfo {author} {\bibfnamefont {F.}~\bibnamefont {Ginelli}},
  \bibinfo {author} {\bibfnamefont {G.}~\bibnamefont {Gr{\'e}goire}}, \ and\
  \bibinfo {author} {\bibfnamefont {F.}~\bibnamefont {Raynaud}},\ }\href@noop
  {} {\bibfield  {journal} {\bibinfo  {journal} {Phys. Rev. E}\ }\textbf
  {\bibinfo {volume} {77}},\ \bibinfo {pages} {046113} (\bibinfo {year}
  {2008}{\natexlab{a}})}\BibitemShut {NoStop}%
\bibitem [{\citenamefont {Mart{\'\i}n-G{\'o}mez}\ \emph
  {et~al.}(2018)\citenamefont {Mart{\'\i}n-G{\'o}mez}, \citenamefont {Levis},
  \citenamefont {D{\'\i}az-Guilera},\ and\ \citenamefont
  {Pagonabarraga}}]{martin2018collective}%
  \BibitemOpen
  \bibfield  {author} {\bibinfo {author} {\bibfnamefont {A.}~\bibnamefont
  {Mart{\'\i}n-G{\'o}mez}}, \bibinfo {author} {\bibfnamefont {D.}~\bibnamefont
  {Levis}}, \bibinfo {author} {\bibfnamefont {A.}~\bibnamefont
  {D{\'\i}az-Guilera}}, \ and\ \bibinfo {author} {\bibfnamefont
  {I.}~\bibnamefont {Pagonabarraga}},\ }\href@noop {} {\bibfield  {journal}
  {\bibinfo  {journal} {Soft Matter}\ }\textbf {\bibinfo {volume} {14}},\
  \bibinfo {pages} {2610} (\bibinfo {year} {2018})}\BibitemShut {NoStop}%
\bibitem [{\citenamefont {van~der Linden}\ \emph {et~al.}(2019)\citenamefont
  {van~der Linden}, \citenamefont {Alexander}, \citenamefont {Aarts},\ and\
  \citenamefont {Dauchot}}]{van2019interrupted}%
  \BibitemOpen
  \bibfield  {author} {\bibinfo {author} {\bibfnamefont {M.~N.}\ \bibnamefont
  {van~der Linden}}, \bibinfo {author} {\bibfnamefont {L.~C.}\ \bibnamefont
  {Alexander}}, \bibinfo {author} {\bibfnamefont {D.~G.}\ \bibnamefont
  {Aarts}}, \ and\ \bibinfo {author} {\bibfnamefont {O.}~\bibnamefont
  {Dauchot}},\ }\href@noop {} {\bibfield  {journal} {\bibinfo  {journal} {Phys.
  Rev. Lett.}\ }\textbf {\bibinfo {volume} {123}},\ \bibinfo {pages} {098001}
  (\bibinfo {year} {2019})}\BibitemShut {NoStop}%
\bibitem [{\citenamefont {Sesé-Sansa}\ \emph {et~al.}(2022)\citenamefont
  {Sesé-Sansa}, \citenamefont {Liao}, \citenamefont {Levis}, \citenamefont
  {Pagonabarraga},\ and\ \citenamefont {Klapp}}]{D2SM00385F}%
  \BibitemOpen
  \bibfield  {author} {\bibinfo {author} {\bibfnamefont {E.}~\bibnamefont
  {Sesé-Sansa}}, \bibinfo {author} {\bibfnamefont {G.-J.}\ \bibnamefont
  {Liao}}, \bibinfo {author} {\bibfnamefont {D.}~\bibnamefont {Levis}},
  \bibinfo {author} {\bibfnamefont {I.}~\bibnamefont {Pagonabarraga}}, \ and\
  \bibinfo {author} {\bibfnamefont {S.~H.~L.}\ \bibnamefont {Klapp}},\
  }\href@noop {} {\bibfield  {journal} {\bibinfo  {journal} {Soft Matter}\
  }\textbf {\bibinfo {volume} {18}},\ \bibinfo {pages} {5388} (\bibinfo {year}
  {2022})}\BibitemShut {NoStop}%
\bibitem [{\citenamefont {Pu}\ \emph {et~al.}(2017)\citenamefont {Pu},
  \citenamefont {Jiang},\ and\ \citenamefont {Hou}}]{pu2017reentrant}%
  \BibitemOpen
  \bibfield  {author} {\bibinfo {author} {\bibfnamefont {M.}~\bibnamefont
  {Pu}}, \bibinfo {author} {\bibfnamefont {H.}~\bibnamefont {Jiang}}, \ and\
  \bibinfo {author} {\bibfnamefont {Z.}~\bibnamefont {Hou}},\ }\href@noop {}
  {\bibfield  {journal} {\bibinfo  {journal} {Soft Matter}\ }\textbf {\bibinfo
  {volume} {13}},\ \bibinfo {pages} {4112} (\bibinfo {year}
  {2017})}\BibitemShut {NoStop}%
\bibitem [{\citenamefont {Knezevic}\ and\ \citenamefont
  {Stark}(2022)}]{knezevic2022collective}%
  \BibitemOpen
  \bibfield  {author} {\bibinfo {author} {\bibfnamefont {M.}~\bibnamefont
  {Knezevic}}\ and\ \bibinfo {author} {\bibfnamefont {H.}~\bibnamefont
  {Stark}},\ }\href@noop {} {\bibfield  {journal} {\bibinfo  {journal} {arXiv
  preprint arXiv:2204.06089}\ } (\bibinfo {year} {2022})}\BibitemShut {NoStop}%
\bibitem [{\citenamefont {Vicsek}\ and\ \citenamefont
  {Zafeiris}(2012)}]{vicsek2012collective}%
  \BibitemOpen
  \bibfield  {author} {\bibinfo {author} {\bibfnamefont {T.}~\bibnamefont
  {Vicsek}}\ and\ \bibinfo {author} {\bibfnamefont {A.}~\bibnamefont
  {Zafeiris}},\ }\href@noop {} {\bibfield  {journal} {\bibinfo  {journal}
  {Phys. Rep.}\ }\textbf {\bibinfo {volume} {517}},\ \bibinfo {pages} {71}
  (\bibinfo {year} {2012})}\BibitemShut {NoStop}%
\bibitem [{\citenamefont {Chat{\'e}}\ \emph
  {et~al.}(2008{\natexlab{b}})\citenamefont {Chat{\'e}}, \citenamefont
  {Ginelli}, \citenamefont {Gr{\'e}goire}, \citenamefont {Peruani},\ and\
  \citenamefont {Raynaud}}]{chate2008modeling}%
  \BibitemOpen
  \bibfield  {author} {\bibinfo {author} {\bibfnamefont {H.}~\bibnamefont
  {Chat{\'e}}}, \bibinfo {author} {\bibfnamefont {F.}~\bibnamefont {Ginelli}},
  \bibinfo {author} {\bibfnamefont {G.}~\bibnamefont {Gr{\'e}goire}}, \bibinfo
  {author} {\bibfnamefont {F.}~\bibnamefont {Peruani}}, \ and\ \bibinfo
  {author} {\bibfnamefont {F.}~\bibnamefont {Raynaud}},\ }\href@noop {}
  {\bibfield  {journal} {\bibinfo  {journal} {Eur. Phys. J. B}\ }\textbf
  {\bibinfo {volume} {64}},\ \bibinfo {pages} {451} (\bibinfo {year}
  {2008}{\natexlab{b}})}\BibitemShut {NoStop}%
\bibitem [{\citenamefont {Cavagna}\ \emph {et~al.}(2015)\citenamefont
  {Cavagna}, \citenamefont {Del~Castello}, \citenamefont {Giardina},
  \citenamefont {Grigera}, \citenamefont {Jelic}, \citenamefont {Melillo},
  \citenamefont {Mora}, \citenamefont {Parisi}, \citenamefont {Silvestri},
  \citenamefont {Viale} \emph {et~al.}}]{cavagna2015flocking}%
  \BibitemOpen
  \bibfield  {author} {\bibinfo {author} {\bibfnamefont {A.}~\bibnamefont
  {Cavagna}}, \bibinfo {author} {\bibfnamefont {L.}~\bibnamefont
  {Del~Castello}}, \bibinfo {author} {\bibfnamefont {I.}~\bibnamefont
  {Giardina}}, \bibinfo {author} {\bibfnamefont {T.}~\bibnamefont {Grigera}},
  \bibinfo {author} {\bibfnamefont {A.}~\bibnamefont {Jelic}}, \bibinfo
  {author} {\bibfnamefont {S.}~\bibnamefont {Melillo}}, \bibinfo {author}
  {\bibfnamefont {T.}~\bibnamefont {Mora}}, \bibinfo {author} {\bibfnamefont
  {L.}~\bibnamefont {Parisi}}, \bibinfo {author} {\bibfnamefont
  {E.}~\bibnamefont {Silvestri}}, \bibinfo {author} {\bibfnamefont
  {M.}~\bibnamefont {Viale}},  \emph {et~al.},\ }\href@noop {} {\bibfield
  {journal} {\bibinfo  {journal} {J. Stat. Phys.}\ }\textbf {\bibinfo {volume}
  {158}},\ \bibinfo {pages} {601} (\bibinfo {year} {2015})}\BibitemShut
  {NoStop}%
\bibitem [{\citenamefont {Liebchen}\ and\ \citenamefont
  {Levis}(2017)}]{liebchen2017collective}%
  \BibitemOpen
  \bibfield  {author} {\bibinfo {author} {\bibfnamefont {B.}~\bibnamefont
  {Liebchen}}\ and\ \bibinfo {author} {\bibfnamefont {D.}~\bibnamefont
  {Levis}},\ }\href@noop {} {\bibfield  {journal} {\bibinfo  {journal} {Phys.
  Rev. Lett.}\ }\textbf {\bibinfo {volume} {119}},\ \bibinfo {pages} {058002}
  (\bibinfo {year} {2017})}\BibitemShut {NoStop}%
\bibitem [{\citenamefont {Cavagna}\ \emph {et~al.}(2018)\citenamefont
  {Cavagna}, \citenamefont {Giardina},\ and\ \citenamefont
  {Grigera}}]{cavagna2018physics}%
  \BibitemOpen
  \bibfield  {author} {\bibinfo {author} {\bibfnamefont {A.}~\bibnamefont
  {Cavagna}}, \bibinfo {author} {\bibfnamefont {I.}~\bibnamefont {Giardina}}, \
  and\ \bibinfo {author} {\bibfnamefont {T.~S.}\ \bibnamefont {Grigera}},\
  }\href@noop {} {\bibfield  {journal} {\bibinfo  {journal} {Phys. Rep.}\
  }\textbf {\bibinfo {volume} {728}},\ \bibinfo {pages} {1} (\bibinfo {year}
  {2018})}\BibitemShut {NoStop}%
\bibitem [{\citenamefont {Palacci}\ \emph {et~al.}(2013)\citenamefont
  {Palacci}, \citenamefont {Sacanna}, \citenamefont {Steinberg}, \citenamefont
  {Pine},\ and\ \citenamefont {Chaikin}}]{palacci2013living}%
  \BibitemOpen
  \bibfield  {author} {\bibinfo {author} {\bibfnamefont {J.}~\bibnamefont
  {Palacci}}, \bibinfo {author} {\bibfnamefont {S.}~\bibnamefont {Sacanna}},
  \bibinfo {author} {\bibfnamefont {A.~P.}\ \bibnamefont {Steinberg}}, \bibinfo
  {author} {\bibfnamefont {D.~J.}\ \bibnamefont {Pine}}, \ and\ \bibinfo
  {author} {\bibfnamefont {P.~M.}\ \bibnamefont {Chaikin}},\ }\href@noop {}
  {\bibfield  {journal} {\bibinfo  {journal} {Sci.}\ }\textbf {\bibinfo
  {volume} {339}},\ \bibinfo {pages} {936} (\bibinfo {year}
  {2013})}\BibitemShut {NoStop}%
\bibitem [{\citenamefont {Fily}\ and\ \citenamefont
  {Marchetti}(2012)}]{fily2012athermal}%
  \BibitemOpen
  \bibfield  {author} {\bibinfo {author} {\bibfnamefont {Y.}~\bibnamefont
  {Fily}}\ and\ \bibinfo {author} {\bibfnamefont {M.~C.}\ \bibnamefont
  {Marchetti}},\ }\href@noop {} {\bibfield  {journal} {\bibinfo  {journal}
  {Phys. Rev. Lett.}\ }\textbf {\bibinfo {volume} {108}},\ \bibinfo {pages}
  {235702} (\bibinfo {year} {2012})}\BibitemShut {NoStop}%
\bibitem [{\citenamefont {Buttinoni}\ \emph {et~al.}(2013)\citenamefont
  {Buttinoni}, \citenamefont {Bialk{\'e}}, \citenamefont {K{\"u}mmel},
  \citenamefont {L{\"o}wen}, \citenamefont {Bechinger},\ and\ \citenamefont
  {Speck}}]{buttinoni2013dynamical}%
  \BibitemOpen
  \bibfield  {author} {\bibinfo {author} {\bibfnamefont {I.}~\bibnamefont
  {Buttinoni}}, \bibinfo {author} {\bibfnamefont {J.}~\bibnamefont
  {Bialk{\'e}}}, \bibinfo {author} {\bibfnamefont {F.}~\bibnamefont
  {K{\"u}mmel}}, \bibinfo {author} {\bibfnamefont {H.}~\bibnamefont
  {L{\"o}wen}}, \bibinfo {author} {\bibfnamefont {C.}~\bibnamefont
  {Bechinger}}, \ and\ \bibinfo {author} {\bibfnamefont {T.}~\bibnamefont
  {Speck}},\ }\href@noop {} {\bibfield  {journal} {\bibinfo  {journal} {Phys.
  Rev. Lett.}\ }\textbf {\bibinfo {volume} {110}},\ \bibinfo {pages} {238301}
  (\bibinfo {year} {2013})}\BibitemShut {NoStop}%
\bibitem [{\citenamefont {Cates}\ and\ \citenamefont
  {Tailleur}(2015)}]{cates2015motility}%
  \BibitemOpen
  \bibfield  {author} {\bibinfo {author} {\bibfnamefont {M.~E.}\ \bibnamefont
  {Cates}}\ and\ \bibinfo {author} {\bibfnamefont {J.}~\bibnamefont
  {Tailleur}},\ }\href@noop {} {\bibfield  {journal} {\bibinfo  {journal}
  {Annu. Rev. Condens. Matter Phys.}\ }\textbf {\bibinfo {volume} {6}},\
  \bibinfo {pages} {219} (\bibinfo {year} {2015})}\BibitemShut {NoStop}%
\bibitem [{\citenamefont {Gonnella}\ \emph {et~al.}(2015)\citenamefont
  {Gonnella}, \citenamefont {Marenduzzo}, \citenamefont {Suma},\ and\
  \citenamefont {Tiribocchi}}]{gonnella2015motility}%
  \BibitemOpen
  \bibfield  {author} {\bibinfo {author} {\bibfnamefont {G.}~\bibnamefont
  {Gonnella}}, \bibinfo {author} {\bibfnamefont {D.}~\bibnamefont
  {Marenduzzo}}, \bibinfo {author} {\bibfnamefont {A.}~\bibnamefont {Suma}}, \
  and\ \bibinfo {author} {\bibfnamefont {A.}~\bibnamefont {Tiribocchi}},\
  }\href@noop {} {\bibfield  {journal} {\bibinfo  {journal} {C. R. Phys.}\
  }\textbf {\bibinfo {volume} {16}},\ \bibinfo {pages} {316} (\bibinfo {year}
  {2015})}\BibitemShut {NoStop}%
\bibitem [{\citenamefont {Marchetti}\ \emph {et~al.}(2016)\citenamefont
  {Marchetti}, \citenamefont {Fily}, \citenamefont {Henkes}, \citenamefont
  {Patch},\ and\ \citenamefont {Yllanes}}]{marchetti2016minimal}%
  \BibitemOpen
  \bibfield  {author} {\bibinfo {author} {\bibfnamefont {M.~C.}\ \bibnamefont
  {Marchetti}}, \bibinfo {author} {\bibfnamefont {Y.}~\bibnamefont {Fily}},
  \bibinfo {author} {\bibfnamefont {S.}~\bibnamefont {Henkes}}, \bibinfo
  {author} {\bibfnamefont {A.}~\bibnamefont {Patch}}, \ and\ \bibinfo {author}
  {\bibfnamefont {D.}~\bibnamefont {Yllanes}},\ }\href@noop {} {\bibfield
  {journal} {\bibinfo  {journal} {Curr. Opin. Colloid Interface Sci.}\ }\textbf
  {\bibinfo {volume} {21}},\ \bibinfo {pages} {34} (\bibinfo {year}
  {2016})}\BibitemShut {NoStop}%
\bibitem [{\citenamefont {Farage}\ \emph {et~al.}(2015)\citenamefont {Farage},
  \citenamefont {Krinninger},\ and\ \citenamefont
  {Brader}}]{farage2015effective}%
  \BibitemOpen
  \bibfield  {author} {\bibinfo {author} {\bibfnamefont {T.~F.}\ \bibnamefont
  {Farage}}, \bibinfo {author} {\bibfnamefont {P.}~\bibnamefont {Krinninger}},
  \ and\ \bibinfo {author} {\bibfnamefont {J.~M.}\ \bibnamefont {Brader}},\
  }\href@noop {} {\bibfield  {journal} {\bibinfo  {journal} {Phys. Rev. E}\
  }\textbf {\bibinfo {volume} {91}},\ \bibinfo {pages} {042310} (\bibinfo
  {year} {2015})}\BibitemShut {NoStop}%
\bibitem [{\citenamefont {Wittmann}\ \emph {et~al.}(2017)\citenamefont
  {Wittmann}, \citenamefont {Maggi}, \citenamefont {Sharma}, \citenamefont
  {Scacchi}, \citenamefont {Brader},\ and\ \citenamefont
  {Marconi}}]{wittmann2017effective}%
  \BibitemOpen
  \bibfield  {author} {\bibinfo {author} {\bibfnamefont {R.}~\bibnamefont
  {Wittmann}}, \bibinfo {author} {\bibfnamefont {C.}~\bibnamefont {Maggi}},
  \bibinfo {author} {\bibfnamefont {A.}~\bibnamefont {Sharma}}, \bibinfo
  {author} {\bibfnamefont {A.}~\bibnamefont {Scacchi}}, \bibinfo {author}
  {\bibfnamefont {J.~M.}\ \bibnamefont {Brader}}, \ and\ \bibinfo {author}
  {\bibfnamefont {U.~M.~B.}\ \bibnamefont {Marconi}},\ }\href@noop {}
  {\bibfield  {journal} {\bibinfo  {journal} {J. Stat. Mech. Theory Exp.}\
  }\textbf {\bibinfo {volume} {2017}},\ \bibinfo {pages} {113207} (\bibinfo
  {year} {2017})}\BibitemShut {NoStop}%
\bibitem [{\citenamefont {Solon}\ \emph {et~al.}(2018)\citenamefont {Solon},
  \citenamefont {Stenhammar}, \citenamefont {Cates}, \citenamefont {Kafri},\
  and\ \citenamefont {Tailleur}}]{solon2018generalized}%
  \BibitemOpen
  \bibfield  {author} {\bibinfo {author} {\bibfnamefont {A.~P.}\ \bibnamefont
  {Solon}}, \bibinfo {author} {\bibfnamefont {J.}~\bibnamefont {Stenhammar}},
  \bibinfo {author} {\bibfnamefont {M.~E.}\ \bibnamefont {Cates}}, \bibinfo
  {author} {\bibfnamefont {Y.}~\bibnamefont {Kafri}}, \ and\ \bibinfo {author}
  {\bibfnamefont {J.}~\bibnamefont {Tailleur}},\ }\href@noop {} {\bibfield
  {journal} {\bibinfo  {journal} {Phys. Rev. E}\ }\textbf {\bibinfo {volume}
  {97}},\ \bibinfo {pages} {020602} (\bibinfo {year} {2018})}\BibitemShut
  {NoStop}%
\bibitem [{\citenamefont {Hermann}\ \emph {et~al.}(2021)\citenamefont
  {Hermann}, \citenamefont {de~las Heras},\ and\ \citenamefont
  {Schmidt}}]{hermann2021phase}%
  \BibitemOpen
  \bibfield  {author} {\bibinfo {author} {\bibfnamefont {S.}~\bibnamefont
  {Hermann}}, \bibinfo {author} {\bibfnamefont {D.}~\bibnamefont {de~las
  Heras}}, \ and\ \bibinfo {author} {\bibfnamefont {M.}~\bibnamefont
  {Schmidt}},\ }\href@noop {} {\bibfield  {journal} {\bibinfo  {journal} {Mol.
  Phys.}\ }\textbf {\bibinfo {volume} {119}},\ \bibinfo {pages} {e1902585}
  (\bibinfo {year} {2021})}\BibitemShut {NoStop}%
\bibitem [{\citenamefont {Speck}\ \emph {et~al.}(2014)\citenamefont {Speck},
  \citenamefont {Bialk{\'e}}, \citenamefont {Menzel},\ and\ \citenamefont
  {L{\"o}wen}}]{speck2014effective}%
  \BibitemOpen
  \bibfield  {author} {\bibinfo {author} {\bibfnamefont {T.}~\bibnamefont
  {Speck}}, \bibinfo {author} {\bibfnamefont {J.}~\bibnamefont {Bialk{\'e}}},
  \bibinfo {author} {\bibfnamefont {A.~M.}\ \bibnamefont {Menzel}}, \ and\
  \bibinfo {author} {\bibfnamefont {H.}~\bibnamefont {L{\"o}wen}},\ }\href@noop
  {} {\bibfield  {journal} {\bibinfo  {journal} {Phys. Rev. Lett.}\ }\textbf
  {\bibinfo {volume} {112}},\ \bibinfo {pages} {218304} (\bibinfo {year}
  {2014})}\BibitemShut {NoStop}%
\bibitem [{\citenamefont {Mandal}\ \emph {et~al.}(2019)\citenamefont {Mandal},
  \citenamefont {Liebchen},\ and\ \citenamefont
  {L{\"o}wen}}]{mandal2019motility}%
  \BibitemOpen
  \bibfield  {author} {\bibinfo {author} {\bibfnamefont {S.}~\bibnamefont
  {Mandal}}, \bibinfo {author} {\bibfnamefont {B.}~\bibnamefont {Liebchen}}, \
  and\ \bibinfo {author} {\bibfnamefont {H.}~\bibnamefont {L{\"o}wen}},\
  }\href@noop {} {\bibfield  {journal} {\bibinfo  {journal} {Phys. Rev. Lett.}\
  }\textbf {\bibinfo {volume} {123}},\ \bibinfo {pages} {228001} (\bibinfo
  {year} {2019})}\BibitemShut {NoStop}%
\bibitem [{\citenamefont {Bialk{\'e}}\ \emph {et~al.}(2015)\citenamefont
  {Bialk{\'e}}, \citenamefont {Siebert}, \citenamefont {L{\"o}wen},\ and\
  \citenamefont {Speck}}]{bialke2015negative}%
  \BibitemOpen
  \bibfield  {author} {\bibinfo {author} {\bibfnamefont {J.}~\bibnamefont
  {Bialk{\'e}}}, \bibinfo {author} {\bibfnamefont {J.~T.}\ \bibnamefont
  {Siebert}}, \bibinfo {author} {\bibfnamefont {H.}~\bibnamefont {L{\"o}wen}},
  \ and\ \bibinfo {author} {\bibfnamefont {T.}~\bibnamefont {Speck}},\
  }\href@noop {} {\bibfield  {journal} {\bibinfo  {journal} {Phys. Rev. Lett.}\
  }\textbf {\bibinfo {volume} {115}},\ \bibinfo {pages} {098301} (\bibinfo
  {year} {2015})}\BibitemShut {NoStop}%
\bibitem [{\citenamefont {Hermann}\ \emph {et~al.}(2019)\citenamefont
  {Hermann}, \citenamefont {de~Las~Heras},\ and\ \citenamefont
  {Schmidt}}]{hermann2019non}%
  \BibitemOpen
  \bibfield  {author} {\bibinfo {author} {\bibfnamefont {S.}~\bibnamefont
  {Hermann}}, \bibinfo {author} {\bibfnamefont {D.}~\bibnamefont
  {de~Las~Heras}}, \ and\ \bibinfo {author} {\bibfnamefont {M.}~\bibnamefont
  {Schmidt}},\ }\href@noop {} {\bibfield  {journal} {\bibinfo  {journal} {Phys.
  Rev. Lett.}\ }\textbf {\bibinfo {volume} {123}},\ \bibinfo {pages} {268002}
  (\bibinfo {year} {2019})}\BibitemShut {NoStop}%
\bibitem [{\citenamefont {Fausti}\ \emph {et~al.}(2021)\citenamefont {Fausti},
  \citenamefont {Tjhung}, \citenamefont {Cates},\ and\ \citenamefont
  {Nardini}}]{fausti2021capillary}%
  \BibitemOpen
  \bibfield  {author} {\bibinfo {author} {\bibfnamefont {G.}~\bibnamefont
  {Fausti}}, \bibinfo {author} {\bibfnamefont {E.}~\bibnamefont {Tjhung}},
  \bibinfo {author} {\bibfnamefont {M.}~\bibnamefont {Cates}}, \ and\ \bibinfo
  {author} {\bibfnamefont {C.}~\bibnamefont {Nardini}},\ }\href@noop {}
  {\bibfield  {journal} {\bibinfo  {journal} {Phys. Rev. Lett.}\ }\textbf
  {\bibinfo {volume} {127}},\ \bibinfo {pages} {068001} (\bibinfo {year}
  {2021})}\BibitemShut {NoStop}%
\bibitem [{\citenamefont {Caprini}\ \emph
  {et~al.}(2020{\natexlab{a}})\citenamefont {Caprini}, \citenamefont
  {Marconi},\ and\ \citenamefont {Puglisi}}]{caprini2020spontaneous}%
  \BibitemOpen
  \bibfield  {author} {\bibinfo {author} {\bibfnamefont {L.}~\bibnamefont
  {Caprini}}, \bibinfo {author} {\bibfnamefont {U.~M.~B.}\ \bibnamefont
  {Marconi}}, \ and\ \bibinfo {author} {\bibfnamefont {A.}~\bibnamefont
  {Puglisi}},\ }\href@noop {} {\bibfield  {journal} {\bibinfo  {journal} {Phys.
  Rev. Lett.}\ }\textbf {\bibinfo {volume} {124}},\ \bibinfo {pages} {078001}
  (\bibinfo {year} {2020}{\natexlab{a}})}\BibitemShut {NoStop}%
\bibitem [{\citenamefont {Caprini}\ \emph
  {et~al.}(2022{\natexlab{a}})\citenamefont {Caprini}, \citenamefont {Gupta},\
  and\ \citenamefont {L{\"o}wen}}]{caprini2022role}%
  \BibitemOpen
  \bibfield  {author} {\bibinfo {author} {\bibfnamefont {L.}~\bibnamefont
  {Caprini}}, \bibinfo {author} {\bibfnamefont {R.~K.}\ \bibnamefont {Gupta}},
  \ and\ \bibinfo {author} {\bibfnamefont {H.}~\bibnamefont {L{\"o}wen}},\
  }\href@noop {} {\bibfield  {journal} {\bibinfo  {journal} {Phys. Chem. Chem.
  Phys.}\ }\textbf {\bibinfo {volume} {24}},\ \bibinfo {pages} {24910}
  (\bibinfo {year} {2022}{\natexlab{a}})}\BibitemShut {NoStop}%
\bibitem [{\citenamefont {Digregorio}\ \emph {et~al.}(2018)\citenamefont
  {Digregorio}, \citenamefont {Levis}, \citenamefont {Suma}, \citenamefont
  {Cugliandolo}, \citenamefont {Gonnella},\ and\ \citenamefont
  {Pagonabarraga}}]{digregorio2018full}%
  \BibitemOpen
  \bibfield  {author} {\bibinfo {author} {\bibfnamefont {P.}~\bibnamefont
  {Digregorio}}, \bibinfo {author} {\bibfnamefont {D.}~\bibnamefont {Levis}},
  \bibinfo {author} {\bibfnamefont {A.}~\bibnamefont {Suma}}, \bibinfo {author}
  {\bibfnamefont {L.~F.}\ \bibnamefont {Cugliandolo}}, \bibinfo {author}
  {\bibfnamefont {G.}~\bibnamefont {Gonnella}}, \ and\ \bibinfo {author}
  {\bibfnamefont {I.}~\bibnamefont {Pagonabarraga}},\ }\href@noop {} {\bibfield
   {journal} {\bibinfo  {journal} {Phys. Rev. Lett.}\ }\textbf {\bibinfo
  {volume} {121}},\ \bibinfo {pages} {098003} (\bibinfo {year}
  {2018})}\BibitemShut {NoStop}%
\bibitem [{\citenamefont {Digregorio}\ \emph {et~al.}(2022)\citenamefont
  {Digregorio}, \citenamefont {Levis}, \citenamefont {Cugliandolo},
  \citenamefont {Gonnella},\ and\ \citenamefont
  {Pagonabarraga}}]{digregorio2022unified}%
  \BibitemOpen
  \bibfield  {author} {\bibinfo {author} {\bibfnamefont {P.}~\bibnamefont
  {Digregorio}}, \bibinfo {author} {\bibfnamefont {D.}~\bibnamefont {Levis}},
  \bibinfo {author} {\bibfnamefont {L.~F.}\ \bibnamefont {Cugliandolo}},
  \bibinfo {author} {\bibfnamefont {G.}~\bibnamefont {Gonnella}}, \ and\
  \bibinfo {author} {\bibfnamefont {I.}~\bibnamefont {Pagonabarraga}},\
  }\href@noop {} {\bibfield  {journal} {\bibinfo  {journal} {Soft Matter}\
  }\textbf {\bibinfo {volume} {18}},\ \bibinfo {pages} {566} (\bibinfo {year}
  {2022})}\BibitemShut {NoStop}%
\bibitem [{\citenamefont {Caporusso}\ \emph {et~al.}(2020)\citenamefont
  {Caporusso}, \citenamefont {Digregorio}, \citenamefont {Levis}, \citenamefont
  {Cugliandolo},\ and\ \citenamefont {Gonnella}}]{caporusso2020motility}%
  \BibitemOpen
  \bibfield  {author} {\bibinfo {author} {\bibfnamefont {C.~B.}\ \bibnamefont
  {Caporusso}}, \bibinfo {author} {\bibfnamefont {P.}~\bibnamefont
  {Digregorio}}, \bibinfo {author} {\bibfnamefont {D.}~\bibnamefont {Levis}},
  \bibinfo {author} {\bibfnamefont {L.~F.}\ \bibnamefont {Cugliandolo}}, \ and\
  \bibinfo {author} {\bibfnamefont {G.}~\bibnamefont {Gonnella}},\ }\href@noop
  {} {\bibfield  {journal} {\bibinfo  {journal} {Phys. Rev. Lett.}\ }\textbf
  {\bibinfo {volume} {125}},\ \bibinfo {pages} {178004} (\bibinfo {year}
  {2020})}\BibitemShut {NoStop}%
\bibitem [{\citenamefont {Shi}\ \emph {et~al.}(2020)\citenamefont {Shi},
  \citenamefont {Fausti}, \citenamefont {Chat{\'e}}, \citenamefont {Nardini},\
  and\ \citenamefont {Solon}}]{shi2020self}%
  \BibitemOpen
  \bibfield  {author} {\bibinfo {author} {\bibfnamefont {X.-q.}\ \bibnamefont
  {Shi}}, \bibinfo {author} {\bibfnamefont {G.}~\bibnamefont {Fausti}},
  \bibinfo {author} {\bibfnamefont {H.}~\bibnamefont {Chat{\'e}}}, \bibinfo
  {author} {\bibfnamefont {C.}~\bibnamefont {Nardini}}, \ and\ \bibinfo
  {author} {\bibfnamefont {A.}~\bibnamefont {Solon}},\ }\href@noop {}
  {\bibfield  {journal} {\bibinfo  {journal} {Phys. Rev. Lett.}\ }\textbf
  {\bibinfo {volume} {125}},\ \bibinfo {pages} {168001} (\bibinfo {year}
  {2020})}\BibitemShut {NoStop}%
\bibitem [{\citenamefont {Caprini}\ \emph
  {et~al.}(2020{\natexlab{b}})\citenamefont {Caprini}, \citenamefont {Marconi},
  \citenamefont {Maggi}, \citenamefont {Paoluzzi},\ and\ \citenamefont
  {Puglisi}}]{caprini2020hidden}%
  \BibitemOpen
  \bibfield  {author} {\bibinfo {author} {\bibfnamefont {L.}~\bibnamefont
  {Caprini}}, \bibinfo {author} {\bibfnamefont {U.~M.~B.}\ \bibnamefont
  {Marconi}}, \bibinfo {author} {\bibfnamefont {C.}~\bibnamefont {Maggi}},
  \bibinfo {author} {\bibfnamefont {M.}~\bibnamefont {Paoluzzi}}, \ and\
  \bibinfo {author} {\bibfnamefont {A.}~\bibnamefont {Puglisi}},\ }\href@noop
  {} {\bibfield  {journal} {\bibinfo  {journal} {Phys. Rev. Research}\ }\textbf
  {\bibinfo {volume} {2}},\ \bibinfo {pages} {023321} (\bibinfo {year}
  {2020}{\natexlab{b}})}\BibitemShut {NoStop}%
\bibitem [{\citenamefont {Gro{\ss}mann}\ \emph {et~al.}(2020)\citenamefont
  {Gro{\ss}mann}, \citenamefont {Aranson},\ and\ \citenamefont
  {Peruani}}]{grossmann2020particle}%
  \BibitemOpen
  \bibfield  {author} {\bibinfo {author} {\bibfnamefont {R.}~\bibnamefont
  {Gro{\ss}mann}}, \bibinfo {author} {\bibfnamefont {I.~S.}\ \bibnamefont
  {Aranson}}, \ and\ \bibinfo {author} {\bibfnamefont {F.}~\bibnamefont
  {Peruani}},\ }\href@noop {} {\bibfield  {journal} {\bibinfo  {journal} {Nat.
  Commun.}\ }\textbf {\bibinfo {volume} {11}},\ \bibinfo {pages} {5365}
  (\bibinfo {year} {2020})}\BibitemShut {NoStop}%
\bibitem [{\citenamefont {Ses{\'e}-Sansa}\ \emph {et~al.}(2021)\citenamefont
  {Ses{\'e}-Sansa}, \citenamefont {Levis},\ and\ \citenamefont
  {Pagonabarraga}}]{sese2021phase}%
  \BibitemOpen
  \bibfield  {author} {\bibinfo {author} {\bibfnamefont {E.}~\bibnamefont
  {Ses{\'e}-Sansa}}, \bibinfo {author} {\bibfnamefont {D.}~\bibnamefont
  {Levis}}, \ and\ \bibinfo {author} {\bibfnamefont {I.}~\bibnamefont
  {Pagonabarraga}},\ }\href@noop {} {\bibfield  {journal} {\bibinfo  {journal}
  {Phys. Rev. E}\ }\textbf {\bibinfo {volume} {104}},\ \bibinfo {pages}
  {054611} (\bibinfo {year} {2021})}\BibitemShut {NoStop}%
\bibitem [{\citenamefont {Lam}\ \emph {et~al.}(2015)\citenamefont {Lam},
  \citenamefont {Schindler},\ and\ \citenamefont {Dauchot}}]{lam2015self}%
  \BibitemOpen
  \bibfield  {author} {\bibinfo {author} {\bibfnamefont {K.-D. N.~T.}\
  \bibnamefont {Lam}}, \bibinfo {author} {\bibfnamefont {M.}~\bibnamefont
  {Schindler}}, \ and\ \bibinfo {author} {\bibfnamefont {O.}~\bibnamefont
  {Dauchot}},\ }\href@noop {} {\bibfield  {journal} {\bibinfo  {journal} {New
  J. Phys.}\ }\textbf {\bibinfo {volume} {17}},\ \bibinfo {pages} {113056}
  (\bibinfo {year} {2015})}\BibitemShut {NoStop}%
\bibitem [{\citenamefont {Binder}\ and\ \citenamefont
  {Virnau}(2021)}]{binder2021phase}%
  \BibitemOpen
  \bibfield  {author} {\bibinfo {author} {\bibfnamefont {K.}~\bibnamefont
  {Binder}}\ and\ \bibinfo {author} {\bibfnamefont {P.}~\bibnamefont
  {Virnau}},\ }\href@noop {} {\bibfield  {journal} {\bibinfo  {journal} {Soft
  Materials}\ }\textbf {\bibinfo {volume} {19}},\ \bibinfo {pages} {263}
  (\bibinfo {year} {2021})}\BibitemShut {NoStop}%
\bibitem [{\citenamefont {Prymidis}\ \emph {et~al.}(2015)\citenamefont
  {Prymidis}, \citenamefont {Sielcken},\ and\ \citenamefont
  {Filion}}]{prymidis2015self}%
  \BibitemOpen
  \bibfield  {author} {\bibinfo {author} {\bibfnamefont {V.}~\bibnamefont
  {Prymidis}}, \bibinfo {author} {\bibfnamefont {H.}~\bibnamefont {Sielcken}},
  \ and\ \bibinfo {author} {\bibfnamefont {L.}~\bibnamefont {Filion}},\
  }\href@noop {} {\bibfield  {journal} {\bibinfo  {journal} {Soft Matter}\
  }\textbf {\bibinfo {volume} {11}},\ \bibinfo {pages} {4158} (\bibinfo {year}
  {2015})}\BibitemShut {NoStop}%
\bibitem [{\citenamefont {Rein}\ and\ \citenamefont
  {Speck}(2016)}]{rein2016applicability}%
  \BibitemOpen
  \bibfield  {author} {\bibinfo {author} {\bibfnamefont {M.}~\bibnamefont
  {Rein}}\ and\ \bibinfo {author} {\bibfnamefont {T.}~\bibnamefont {Speck}},\
  }\href@noop {} {\bibfield  {journal} {\bibinfo  {journal} {Eur. Phys. J. E}\
  }\textbf {\bibinfo {volume} {39}},\ \bibinfo {pages} {84} (\bibinfo {year}
  {2016})}\BibitemShut {NoStop}%
\bibitem [{\citenamefont {Paliwal}\ \emph {et~al.}(2017)\citenamefont
  {Paliwal}, \citenamefont {Prymidis}, \citenamefont {Filion},\ and\
  \citenamefont {Dijkstra}}]{paliwal2017non}%
  \BibitemOpen
  \bibfield  {author} {\bibinfo {author} {\bibfnamefont {S.}~\bibnamefont
  {Paliwal}}, \bibinfo {author} {\bibfnamefont {V.}~\bibnamefont {Prymidis}},
  \bibinfo {author} {\bibfnamefont {L.}~\bibnamefont {Filion}}, \ and\ \bibinfo
  {author} {\bibfnamefont {M.}~\bibnamefont {Dijkstra}},\ }\href@noop {}
  {\bibfield  {journal} {\bibinfo  {journal} {J. Chem. Phys.}\ }\textbf
  {\bibinfo {volume} {147}},\ \bibinfo {pages} {084902} (\bibinfo {year}
  {2017})}\BibitemShut {NoStop}%
\bibitem [{\citenamefont {W{\"a}chtler}\ \emph {et~al.}(2016)\citenamefont
  {W{\"a}chtler}, \citenamefont {Kogler},\ and\ \citenamefont
  {Klapp}}]{wachtler2016lane}%
  \BibitemOpen
  \bibfield  {author} {\bibinfo {author} {\bibfnamefont {C.}~\bibnamefont
  {W{\"a}chtler}}, \bibinfo {author} {\bibfnamefont {F.}~\bibnamefont
  {Kogler}}, \ and\ \bibinfo {author} {\bibfnamefont {S.}~\bibnamefont
  {Klapp}},\ }\href@noop {} {\bibfield  {journal} {\bibinfo  {journal} {Phys.
  Rev. E}\ }\textbf {\bibinfo {volume} {94}},\ \bibinfo {pages} {052603}
  (\bibinfo {year} {2016})}\BibitemShut {NoStop}%
\bibitem [{\citenamefont {Mallory}\ \emph {et~al.}(2017)\citenamefont
  {Mallory}, \citenamefont {Alarcon}, \citenamefont {Cacciuto},\ and\
  \citenamefont {Valeriani}}]{mallory2017self}%
  \BibitemOpen
  \bibfield  {author} {\bibinfo {author} {\bibfnamefont {S.}~\bibnamefont
  {Mallory}}, \bibinfo {author} {\bibfnamefont {F.}~\bibnamefont {Alarcon}},
  \bibinfo {author} {\bibfnamefont {A.}~\bibnamefont {Cacciuto}}, \ and\
  \bibinfo {author} {\bibfnamefont {C.}~\bibnamefont {Valeriani}},\ }\href@noop
  {} {\bibfield  {journal} {\bibinfo  {journal} {New J. Phys.}\ }\textbf
  {\bibinfo {volume} {19}},\ \bibinfo {pages} {125014} (\bibinfo {year}
  {2017})}\BibitemShut {NoStop}%
\bibitem [{\citenamefont {Mani}\ and\ \citenamefont
  {L{\"o}wen}(2015)}]{mani2015effect}%
  \BibitemOpen
  \bibfield  {author} {\bibinfo {author} {\bibfnamefont {E.}~\bibnamefont
  {Mani}}\ and\ \bibinfo {author} {\bibfnamefont {H.}~\bibnamefont
  {L{\"o}wen}},\ }\href@noop {} {\bibfield  {journal} {\bibinfo  {journal}
  {Phys. Rev. E}\ }\textbf {\bibinfo {volume} {92}},\ \bibinfo {pages} {032301}
  (\bibinfo {year} {2015})}\BibitemShut {NoStop}%
\bibitem [{\citenamefont {Alarc{\'o}n}\ \emph {et~al.}(2017)\citenamefont
  {Alarc{\'o}n}, \citenamefont {Valeriani},\ and\ \citenamefont
  {Pagonabarraga}}]{alarcon2017morphology}%
  \BibitemOpen
  \bibfield  {author} {\bibinfo {author} {\bibfnamefont {F.}~\bibnamefont
  {Alarc{\'o}n}}, \bibinfo {author} {\bibfnamefont {C.}~\bibnamefont
  {Valeriani}}, \ and\ \bibinfo {author} {\bibfnamefont {I.}~\bibnamefont
  {Pagonabarraga}},\ }\href@noop {} {\bibfield  {journal} {\bibinfo  {journal}
  {Soft Matter}\ }\textbf {\bibinfo {volume} {13}},\ \bibinfo {pages} {814}
  (\bibinfo {year} {2017})}\BibitemShut {NoStop}%
\bibitem [{\citenamefont {Navarro}\ and\ \citenamefont
  {Fielding}(2015)}]{navarro2015clustering}%
  \BibitemOpen
  \bibfield  {author} {\bibinfo {author} {\bibfnamefont {R.~M.}\ \bibnamefont
  {Navarro}}\ and\ \bibinfo {author} {\bibfnamefont {S.~M.}\ \bibnamefont
  {Fielding}},\ }\href@noop {} {\bibfield  {journal} {\bibinfo  {journal} {Soft
  Matter}\ }\textbf {\bibinfo {volume} {11}},\ \bibinfo {pages} {7525}
  (\bibinfo {year} {2015})}\BibitemShut {NoStop}%
\bibitem [{\citenamefont {Redner}\ \emph
  {et~al.}(2013{\natexlab{a}})\citenamefont {Redner}, \citenamefont
  {Baskaran},\ and\ \citenamefont {Hagan}}]{redner2013reentrant}%
  \BibitemOpen
  \bibfield  {author} {\bibinfo {author} {\bibfnamefont {G.}~\bibnamefont
  {Redner}}, \bibinfo {author} {\bibfnamefont {A.}~\bibnamefont {Baskaran}}, \
  and\ \bibinfo {author} {\bibfnamefont {M.}~\bibnamefont {Hagan}},\
  }\href@noop {} {\bibfield  {journal} {\bibinfo  {journal} {Phys. Rev. E}\
  }\textbf {\bibinfo {volume} {88}},\ \bibinfo {pages} {012305} (\bibinfo
  {year} {2013}{\natexlab{a}})}\BibitemShut {NoStop}%
\bibitem [{\citenamefont {Prymidis}\ \emph {et~al.}(2016)\citenamefont
  {Prymidis}, \citenamefont {Paliwal}, \citenamefont {Dijkstra},\ and\
  \citenamefont {Filion}}]{prymidis2016vapour}%
  \BibitemOpen
  \bibfield  {author} {\bibinfo {author} {\bibfnamefont {V.}~\bibnamefont
  {Prymidis}}, \bibinfo {author} {\bibfnamefont {S.}~\bibnamefont {Paliwal}},
  \bibinfo {author} {\bibfnamefont {M.}~\bibnamefont {Dijkstra}}, \ and\
  \bibinfo {author} {\bibfnamefont {L.}~\bibnamefont {Filion}},\ }\href@noop {}
  {\bibfield  {journal} {\bibinfo  {journal} {J. Chem. Phys.}\ }\textbf
  {\bibinfo {volume} {145}},\ \bibinfo {pages} {124904} (\bibinfo {year}
  {2016})}\BibitemShut {NoStop}%
\bibitem [{\citenamefont {Hrishikesh}\ and\ \citenamefont
  {Mani}(2022)}]{hrishikesh2022collective}%
  \BibitemOpen
  \bibfield  {author} {\bibinfo {author} {\bibfnamefont {B.}~\bibnamefont
  {Hrishikesh}}\ and\ \bibinfo {author} {\bibfnamefont {E.}~\bibnamefont
  {Mani}},\ }\href@noop {} {\bibfield  {journal} {\bibinfo  {journal} {Phys.
  Chem. Chem. Phys.}\ }\textbf {\bibinfo {volume} {24}},\ \bibinfo {pages}
  {19792} (\bibinfo {year} {2022})}\BibitemShut {NoStop}%
\bibitem [{\citenamefont {Shaebani}\ \emph {et~al.}(2020)\citenamefont
  {Shaebani}, \citenamefont {Wysocki}, \citenamefont {Winkler}, \citenamefont
  {Gompper},\ and\ \citenamefont {Rieger}}]{shaebani2020computational}%
  \BibitemOpen
  \bibfield  {author} {\bibinfo {author} {\bibfnamefont {M.~R.}\ \bibnamefont
  {Shaebani}}, \bibinfo {author} {\bibfnamefont {A.}~\bibnamefont {Wysocki}},
  \bibinfo {author} {\bibfnamefont {R.~G.}\ \bibnamefont {Winkler}}, \bibinfo
  {author} {\bibfnamefont {G.}~\bibnamefont {Gompper}}, \ and\ \bibinfo
  {author} {\bibfnamefont {H.}~\bibnamefont {Rieger}},\ }\href@noop {}
  {\bibfield  {journal} {\bibinfo  {journal} {Nat. Rev. Phys.}\ }\textbf
  {\bibinfo {volume} {2}},\ \bibinfo {pages} {181} (\bibinfo {year}
  {2020})}\BibitemShut {NoStop}%
\bibitem [{\citenamefont {Redner}\ \emph
  {et~al.}(2013{\natexlab{b}})\citenamefont {Redner}, \citenamefont {Hagan},\
  and\ \citenamefont {Baskaran}}]{redner2013structure}%
  \BibitemOpen
  \bibfield  {author} {\bibinfo {author} {\bibfnamefont {G.~S.}\ \bibnamefont
  {Redner}}, \bibinfo {author} {\bibfnamefont {M.~F.}\ \bibnamefont {Hagan}}, \
  and\ \bibinfo {author} {\bibfnamefont {A.}~\bibnamefont {Baskaran}},\
  }\href@noop {} {\bibfield  {journal} {\bibinfo  {journal} {Phys. Rev. Lett.}\
  }\textbf {\bibinfo {volume} {110}},\ \bibinfo {pages} {055701} (\bibinfo
  {year} {2013}{\natexlab{b}})}\BibitemShut {NoStop}%
\bibitem [{\citenamefont {Mognetti}\ \emph {et~al.}(2013)\citenamefont
  {Mognetti}, \citenamefont {{\v{S}}ari{\'c}}, \citenamefont
  {Angioletti-Uberti}, \citenamefont {Cacciuto}, \citenamefont {Valeriani},\
  and\ \citenamefont {Frenkel}}]{mognetti2013living}%
  \BibitemOpen
  \bibfield  {author} {\bibinfo {author} {\bibfnamefont {B.~M.}\ \bibnamefont
  {Mognetti}}, \bibinfo {author} {\bibfnamefont {A.}~\bibnamefont
  {{\v{S}}ari{\'c}}}, \bibinfo {author} {\bibfnamefont {S.}~\bibnamefont
  {Angioletti-Uberti}}, \bibinfo {author} {\bibfnamefont {A.}~\bibnamefont
  {Cacciuto}}, \bibinfo {author} {\bibfnamefont {C.}~\bibnamefont {Valeriani}},
  \ and\ \bibinfo {author} {\bibfnamefont {D.}~\bibnamefont {Frenkel}},\
  }\href@noop {} {\bibfield  {journal} {\bibinfo  {journal} {Phys. Rev. Lett.}\
  }\textbf {\bibinfo {volume} {111}},\ \bibinfo {pages} {245702} (\bibinfo
  {year} {2013})}\BibitemShut {NoStop}%
\bibitem [{\citenamefont {Cavagna}\ \emph {et~al.}(2022)\citenamefont
  {Cavagna}, \citenamefont {Culla}, \citenamefont {Feng}, \citenamefont
  {Giardina}, \citenamefont {Grigera}, \citenamefont {Kion-Crosby},
  \citenamefont {Melillo}, \citenamefont {Pisegna}, \citenamefont
  {Postiglione},\ and\ \citenamefont {Villegas}}]{cavagna2022marginal}%
  \BibitemOpen
  \bibfield  {author} {\bibinfo {author} {\bibfnamefont {A.}~\bibnamefont
  {Cavagna}}, \bibinfo {author} {\bibfnamefont {A.}~\bibnamefont {Culla}},
  \bibinfo {author} {\bibfnamefont {X.}~\bibnamefont {Feng}}, \bibinfo {author}
  {\bibfnamefont {I.}~\bibnamefont {Giardina}}, \bibinfo {author}
  {\bibfnamefont {T.~S.}\ \bibnamefont {Grigera}}, \bibinfo {author}
  {\bibfnamefont {W.}~\bibnamefont {Kion-Crosby}}, \bibinfo {author}
  {\bibfnamefont {S.}~\bibnamefont {Melillo}}, \bibinfo {author} {\bibfnamefont
  {G.}~\bibnamefont {Pisegna}}, \bibinfo {author} {\bibfnamefont
  {L.}~\bibnamefont {Postiglione}}, \ and\ \bibinfo {author} {\bibfnamefont
  {P.}~\bibnamefont {Villegas}},\ }\href@noop {} {\bibfield  {journal}
  {\bibinfo  {journal} {Nat. Commun.}\ }\textbf {\bibinfo {volume} {13}},\
  \bibinfo {pages} {2315} (\bibinfo {year} {2022})}\BibitemShut {NoStop}%
\bibitem [{\citenamefont {Maggi}\ \emph {et~al.}(2015)\citenamefont {Maggi},
  \citenamefont {Marconi}, \citenamefont {Gnan},\ and\ \citenamefont
  {Di~Leonardo}}]{maggi2015multidimensional}%
  \BibitemOpen
  \bibfield  {author} {\bibinfo {author} {\bibfnamefont {C.}~\bibnamefont
  {Maggi}}, \bibinfo {author} {\bibfnamefont {U.~M.~B.}\ \bibnamefont
  {Marconi}}, \bibinfo {author} {\bibfnamefont {N.}~\bibnamefont {Gnan}}, \
  and\ \bibinfo {author} {\bibfnamefont {R.}~\bibnamefont {Di~Leonardo}},\
  }\href@noop {} {\bibfield  {journal} {\bibinfo  {journal} {Sci. Rep.}\
  }\textbf {\bibinfo {volume} {5}},\ \bibinfo {pages} {10742} (\bibinfo {year}
  {2015})}\BibitemShut {NoStop}%
\bibitem [{\citenamefont {Mandal}\ \emph {et~al.}(2017)\citenamefont {Mandal},
  \citenamefont {Klymko},\ and\ \citenamefont {DeWeese}}]{mandal2017entropy}%
  \BibitemOpen
  \bibfield  {author} {\bibinfo {author} {\bibfnamefont {D.}~\bibnamefont
  {Mandal}}, \bibinfo {author} {\bibfnamefont {K.}~\bibnamefont {Klymko}}, \
  and\ \bibinfo {author} {\bibfnamefont {M.~R.}\ \bibnamefont {DeWeese}},\
  }\href@noop {} {\bibfield  {journal} {\bibinfo  {journal} {Phys. Rev. Lett.}\
  }\textbf {\bibinfo {volume} {119}},\ \bibinfo {pages} {258001} (\bibinfo
  {year} {2017})}\BibitemShut {NoStop}%
\bibitem [{\citenamefont {Berthier}\ \emph {et~al.}(2019)\citenamefont
  {Berthier}, \citenamefont {Flenner},\ and\ \citenamefont
  {Szamel}}]{berthier2019glassy}%
  \BibitemOpen
  \bibfield  {author} {\bibinfo {author} {\bibfnamefont {L.}~\bibnamefont
  {Berthier}}, \bibinfo {author} {\bibfnamefont {E.}~\bibnamefont {Flenner}}, \
  and\ \bibinfo {author} {\bibfnamefont {G.}~\bibnamefont {Szamel}},\
  }\href@noop {} {\bibfield  {journal} {\bibinfo  {journal} {J. Chem. Phys.}\
  }\textbf {\bibinfo {volume} {150}},\ \bibinfo {pages} {200901} (\bibinfo
  {year} {2019})}\BibitemShut {NoStop}%
\bibitem [{\citenamefont {Martin}\ \emph {et~al.}(2021)\citenamefont {Martin},
  \citenamefont {O'Byrne}, \citenamefont {Cates}, \citenamefont {Fodor},
  \citenamefont {Nardini}, \citenamefont {Tailleur},\ and\ \citenamefont {van
  Wijland}}]{martin2021statistical}%
  \BibitemOpen
  \bibfield  {author} {\bibinfo {author} {\bibfnamefont {D.}~\bibnamefont
  {Martin}}, \bibinfo {author} {\bibfnamefont {J.}~\bibnamefont {O'Byrne}},
  \bibinfo {author} {\bibfnamefont {M.~E.}\ \bibnamefont {Cates}}, \bibinfo
  {author} {\bibfnamefont {{\'E}.}~\bibnamefont {Fodor}}, \bibinfo {author}
  {\bibfnamefont {C.}~\bibnamefont {Nardini}}, \bibinfo {author} {\bibfnamefont
  {J.}~\bibnamefont {Tailleur}}, \ and\ \bibinfo {author} {\bibfnamefont
  {F.}~\bibnamefont {van Wijland}},\ }\href@noop {} {\bibfield  {journal}
  {\bibinfo  {journal} {Phys. Rev. E}\ }\textbf {\bibinfo {volume} {103}},\
  \bibinfo {pages} {032607} (\bibinfo {year} {2021})}\BibitemShut {NoStop}%
\bibitem [{\citenamefont {Caprini}\ \emph
  {et~al.}(2022{\natexlab{b}})\citenamefont {Caprini}, \citenamefont
  {Sprenger}, \citenamefont {L{\"o}wen},\ and\ \citenamefont
  {Wittmann}}]{caprini2022parental}%
  \BibitemOpen
  \bibfield  {author} {\bibinfo {author} {\bibfnamefont {L.}~\bibnamefont
  {Caprini}}, \bibinfo {author} {\bibfnamefont {A.~R.}\ \bibnamefont
  {Sprenger}}, \bibinfo {author} {\bibfnamefont {H.}~\bibnamefont {L{\"o}wen}},
  \ and\ \bibinfo {author} {\bibfnamefont {R.}~\bibnamefont {Wittmann}},\
  }\href@noop {} {\bibfield  {journal} {\bibinfo  {journal} {J. Chem. Phys.}\
  }\textbf {\bibinfo {volume} {156}},\ \bibinfo {pages} {071102} (\bibinfo
  {year} {2022}{\natexlab{b}})}\BibitemShut {NoStop}%
\bibitem [{\citenamefont {Fodor}\ \emph {et~al.}(2016)\citenamefont {Fodor},
  \citenamefont {Nardini}, \citenamefont {Cates}, \citenamefont {Tailleur},
  \citenamefont {Visco},\ and\ \citenamefont {van Wijland}}]{fodor2016far}%
  \BibitemOpen
  \bibfield  {author} {\bibinfo {author} {\bibfnamefont {E.}~\bibnamefont
  {Fodor}}, \bibinfo {author} {\bibfnamefont {C.}~\bibnamefont {Nardini}},
  \bibinfo {author} {\bibfnamefont {M.~E.}\ \bibnamefont {Cates}}, \bibinfo
  {author} {\bibfnamefont {J.}~\bibnamefont {Tailleur}}, \bibinfo {author}
  {\bibfnamefont {P.}~\bibnamefont {Visco}}, \ and\ \bibinfo {author}
  {\bibfnamefont {F.}~\bibnamefont {van Wijland}},\ }\href@noop {} {\bibfield
  {journal} {\bibinfo  {journal} {Phys. Rev. Lett.}\ }\textbf {\bibinfo
  {volume} {117}},\ \bibinfo {pages} {038103} (\bibinfo {year}
  {2016})}\BibitemShut {NoStop}%
\bibitem [{\citenamefont {Caprini}\ and\ \citenamefont
  {Marconi}(2018)}]{caprini2018active}%
  \BibitemOpen
  \bibfield  {author} {\bibinfo {author} {\bibfnamefont {L.}~\bibnamefont
  {Caprini}}\ and\ \bibinfo {author} {\bibfnamefont {U.~M.~B.}\ \bibnamefont
  {Marconi}},\ }\href@noop {} {\bibfield  {journal} {\bibinfo  {journal} {Soft
  Matter}\ }\textbf {\bibinfo {volume} {14}},\ \bibinfo {pages} {9044}
  (\bibinfo {year} {2018})}\BibitemShut {NoStop}%
\bibitem [{\citenamefont {Das}\ \emph {et~al.}(2018)\citenamefont {Das},
  \citenamefont {Gompper},\ and\ \citenamefont {Winkler}}]{das2018confined}%
  \BibitemOpen
  \bibfield  {author} {\bibinfo {author} {\bibfnamefont {S.}~\bibnamefont
  {Das}}, \bibinfo {author} {\bibfnamefont {G.}~\bibnamefont {Gompper}}, \ and\
  \bibinfo {author} {\bibfnamefont {R.~G.}\ \bibnamefont {Winkler}},\
  }\href@noop {} {\bibfield  {journal} {\bibinfo  {journal} {New J. Phys.}\
  }\textbf {\bibinfo {volume} {20}},\ \bibinfo {pages} {015001} (\bibinfo
  {year} {2018})}\BibitemShut {NoStop}%
\bibitem [{\citenamefont {Dabelow}\ \emph {et~al.}(2019)\citenamefont
  {Dabelow}, \citenamefont {Bo},\ and\ \citenamefont
  {Eichhorn}}]{dabelow2019irreversibility}%
  \BibitemOpen
  \bibfield  {author} {\bibinfo {author} {\bibfnamefont {L.}~\bibnamefont
  {Dabelow}}, \bibinfo {author} {\bibfnamefont {S.}~\bibnamefont {Bo}}, \ and\
  \bibinfo {author} {\bibfnamefont {R.}~\bibnamefont {Eichhorn}},\ }\href@noop
  {} {\bibfield  {journal} {\bibinfo  {journal} {Phys. Rev. X}\ }\textbf
  {\bibinfo {volume} {9}},\ \bibinfo {pages} {021009} (\bibinfo {year}
  {2019})}\BibitemShut {NoStop}%
\bibitem [{\citenamefont {Caprini}\ and\ \citenamefont {Marini
  Bettolo~Marconi}(2020)}]{caprini2020activehigh}%
  \BibitemOpen
  \bibfield  {author} {\bibinfo {author} {\bibfnamefont {L.}~\bibnamefont
  {Caprini}}\ and\ \bibinfo {author} {\bibfnamefont {U.}~\bibnamefont {Marini
  Bettolo~Marconi}},\ }\href@noop {} {\bibfield  {journal} {\bibinfo  {journal}
  {J. Chem. Phys.}\ }\textbf {\bibinfo {volume} {153}},\ \bibinfo {pages}
  {184901} (\bibinfo {year} {2020})}\BibitemShut {NoStop}%
\bibitem [{\citenamefont {Attanasi}\ \emph {et~al.}(2014)\citenamefont
  {Attanasi}, \citenamefont {Cavagna}, \citenamefont {Del~Castello},
  \citenamefont {Giardina}, \citenamefont {Melillo}, \citenamefont {Parisi},
  \citenamefont {Pohl}, \citenamefont {Rossaro}, \citenamefont {Shen},
  \citenamefont {Silvestri} \emph {et~al.}}]{attanasi2014collective}%
  \BibitemOpen
  \bibfield  {author} {\bibinfo {author} {\bibfnamefont {A.}~\bibnamefont
  {Attanasi}}, \bibinfo {author} {\bibfnamefont {A.}~\bibnamefont {Cavagna}},
  \bibinfo {author} {\bibfnamefont {L.}~\bibnamefont {Del~Castello}}, \bibinfo
  {author} {\bibfnamefont {I.}~\bibnamefont {Giardina}}, \bibinfo {author}
  {\bibfnamefont {S.}~\bibnamefont {Melillo}}, \bibinfo {author} {\bibfnamefont
  {L.}~\bibnamefont {Parisi}}, \bibinfo {author} {\bibfnamefont
  {O.}~\bibnamefont {Pohl}}, \bibinfo {author} {\bibfnamefont {B.}~\bibnamefont
  {Rossaro}}, \bibinfo {author} {\bibfnamefont {E.}~\bibnamefont {Shen}},
  \bibinfo {author} {\bibfnamefont {E.}~\bibnamefont {Silvestri}},  \emph
  {et~al.},\ }\href@noop {} {\bibfield  {journal} {\bibinfo  {journal} {PLoS
  computational biology}\ }\textbf {\bibinfo {volume} {10}},\ \bibinfo {pages}
  {e1003697} (\bibinfo {year} {2014})}\BibitemShut {NoStop}%
\bibitem [{\citenamefont {Dombrovski}\ \emph {et~al.}(2017)\citenamefont
  {Dombrovski}, \citenamefont {Poussard}, \citenamefont {Moalem}, \citenamefont
  {Kmecova}, \citenamefont {Hogan}, \citenamefont {Schott}, \citenamefont
  {Vaccari}, \citenamefont {Acton},\ and\ \citenamefont
  {Condron}}]{dombrovski2017cooperative}%
  \BibitemOpen
  \bibfield  {author} {\bibinfo {author} {\bibfnamefont {M.}~\bibnamefont
  {Dombrovski}}, \bibinfo {author} {\bibfnamefont {L.}~\bibnamefont
  {Poussard}}, \bibinfo {author} {\bibfnamefont {K.}~\bibnamefont {Moalem}},
  \bibinfo {author} {\bibfnamefont {L.}~\bibnamefont {Kmecova}}, \bibinfo
  {author} {\bibfnamefont {N.}~\bibnamefont {Hogan}}, \bibinfo {author}
  {\bibfnamefont {E.}~\bibnamefont {Schott}}, \bibinfo {author} {\bibfnamefont
  {A.}~\bibnamefont {Vaccari}}, \bibinfo {author} {\bibfnamefont
  {S.}~\bibnamefont {Acton}}, \ and\ \bibinfo {author} {\bibfnamefont
  {B.}~\bibnamefont {Condron}},\ }\href@noop {} {\bibfield  {journal} {\bibinfo
   {journal} {Curr. Biol.}\ }\textbf {\bibinfo {volume} {27}},\ \bibinfo
  {pages} {2821} (\bibinfo {year} {2017})}\BibitemShut {NoStop}%
\bibitem [{\citenamefont {Henkes}\ \emph {et~al.}(2020)\citenamefont {Henkes},
  \citenamefont {Kostanjevec}, \citenamefont {Collinson}, \citenamefont
  {Sknepnek},\ and\ \citenamefont {Bertin}}]{henkes2020dense}%
  \BibitemOpen
  \bibfield  {author} {\bibinfo {author} {\bibfnamefont {S.}~\bibnamefont
  {Henkes}}, \bibinfo {author} {\bibfnamefont {K.}~\bibnamefont {Kostanjevec}},
  \bibinfo {author} {\bibfnamefont {J.~M.}\ \bibnamefont {Collinson}}, \bibinfo
  {author} {\bibfnamefont {R.}~\bibnamefont {Sknepnek}}, \ and\ \bibinfo
  {author} {\bibfnamefont {E.}~\bibnamefont {Bertin}},\ }\href@noop {}
  {\bibfield  {journal} {\bibinfo  {journal} {Nat. Commun.}\ }\textbf {\bibinfo
  {volume} {11}},\ \bibinfo {pages} {1405} (\bibinfo {year}
  {2020})}\BibitemShut {NoStop}%
\bibitem [{\citenamefont {Ginot}\ \emph {et~al.}(2018)\citenamefont {Ginot},
  \citenamefont {Theurkauff}, \citenamefont {Detcheverry}, \citenamefont
  {Ybert},\ and\ \citenamefont {Cottin-Bizonne}}]{ginot2018aggregation}%
  \BibitemOpen
  \bibfield  {author} {\bibinfo {author} {\bibfnamefont {F.}~\bibnamefont
  {Ginot}}, \bibinfo {author} {\bibfnamefont {I.}~\bibnamefont {Theurkauff}},
  \bibinfo {author} {\bibfnamefont {F.}~\bibnamefont {Detcheverry}}, \bibinfo
  {author} {\bibfnamefont {C.}~\bibnamefont {Ybert}}, \ and\ \bibinfo {author}
  {\bibfnamefont {C.}~\bibnamefont {Cottin-Bizonne}},\ }\href@noop {}
  {\bibfield  {journal} {\bibinfo  {journal} {Nat. Commun.}\ }\textbf {\bibinfo
  {volume} {9}},\ \bibinfo {pages} {696} (\bibinfo {year} {2018})}\BibitemShut
  {NoStop}%
\end{thebibliography}%

\end{document}